\documentclass{aa}  
\usepackage{graphicx}
\usepackage[colorlinks=true,citecolor=blue]{hyperref}
\usepackage{multirow}

\usepackage{wasysym}

\newcommand{\dd}[0]{\mathrm{d}}

\usepackage{xcolor}

\usepackage{comment}
\usepackage{float}
\usepackage{dblfloatfix}
\usepackage{multicol,multirow}

\defcitealias{Valles-Perez_2023}{VP23}

\usepackage{txfonts}

\begin{document}

   \title{{The eventful life journey of galaxy clusters}}
   \subtitle{I. Impact of DM halo and ICM properties on their full assembly histories}

   \titlerunning{Impact of DM halo and ICM properties on their MAHs}

   \author{David Vallés-Pérez
          \inst{1}\fnmsep\thanks{\email{david.valles-perez@uv.es}.}
          \and
          Susana Planelles\inst{1,2}
          \and 
          Vicent Quilis\inst{1,2}
          }

   \institute{Departament d’Astronomia i Astrofísica, Universitat de València, E-46100 Burjassot (València), Spain
              \and
              Observatori Astronòmic, Universitat de València, E-46980 Paterna (València), Spain}

   \date{\today}

  \abstract
   {Galaxy clusters assemble over gigayears in a very anisotropic environment, which causes a remarkable diversity in their mass assembly histories (MAH).}
   {In this work, we have aimed to understand how the present-day properties of the dark matter halo and the intracluster medium are related to the whole evolution of these structures.}
   {To this end, we analysed a $\Lambda$CDM hydrodynamical+$N$-Body simulation of a $(100 \, h^{-1} \mathrm{Mpc})^3$ volume, containing over 30 clusters and 300 groups, and looked at the individual and the stacked MAHs (determined from complete merger trees) in relation to properties of the DM haloes and the ICM at fixed cosmic time ({indicators of assembly state}).}
   {The ensemble MAHs are well separated when stacked in bins of these indicators, yielding clear dependencies of evolutionary properties (such as formation redshift) on fix-time halo properties. Additionally, we find that different indicators are informative about distinct epochs of accretion. Finally, by summarising the complex MAH diversity with two parameters, we describe how different indicators bring complementary information in different directions of this biparametric space. Overall, halo spin and a combined indicator appear to be the ones encoding the most information about the MAH.}
   {The results shown here add up to the idea that the dynamical state of cosmic structures is a multifaceted concept, and warn that single indicators are uncapable of capturing the whole complexity of the process. This work sheds light on the nature of this characterisation by untangling precisely \textit{when} and \textit{how} several indicators are informative about. In turn, this can provide clues to better constrain the MAH of observed structures.}

   \keywords{galaxies: clusters: general -- galaxies: clusters: intracluster medium -- galaxies:groups:general -- large-scale structure of Universe -- methods: numerical -- methods: statistical}

   \maketitle

\section{Introduction}
\label{s:intro}

Galaxy clusters and groups, spanning a mass range from $10^{13} M_\odot$ to a few times $10^{15} M_\odot$, form through the most energetic (up to ${\sim}~10^{61-65} \, \mathrm{erg}$) and long-lasting (on timescales comparable to the Hubble time) events in the Universe \citep{Kravtsov_2012, Planelles_2015}. Within this process, dark matter (DM) and baryons collapse in an increasingly inhomogeneous and anisotropic environment, due to the emergence of an intricate web of filaments, walls and voids \citep{White_1987, Bond_1996}. Additionally, the process is hierarchical, and thus the mass growth of clusters is contributed both by accretion of smooth material from the cosmic web, and from a more intermittent and violent component associated to mergers of similarly-sized objects \citep{Gott_1975, White_1978}. All this complexity is imprinted in the high diversity of mass assembly histories (MAHs), where, even at fixed present-day mass, there is a very considerable scatter in $M(z)/M_0$ \citep[e.g.][]{Maulbetsch_2007, Zhao_2009, Valles-Perez_2020}.

Characterising the formation history of galaxy clusters and groups is of utmost importance, since it has been linked to a plethora of properties and phenomena of the DM halo and of the intracluster medium (ICM). Regarding the latter, these include, for instance, turbulence \citep{Dolag_2005, Lau_2009, Vazza_2017, Valles-Perez_2021} leading to mass bias \citep{Lau_2013, Nelson_2014, Vazza_2018, Angelinelli_2020, Bennett_2022}, self-similarity or lack thereof, either when looking at integrated properties \citep{McCarthy_2008, Poole_2008, Planelles_2009, Chen_2019} or at radial profiles \citep{Lau_2015}, ram-pressure stripping of infalling satellites \citep{Lourenco_2023}, or gas clumping \citep{Nagai_2011, Vazza_2013, Eckert_2015, Angelinelli_2021}. As per properties of the DM halo, the splashback radius \citep{More_2015, Diemer_2017, Pizzardo_2024} or its shape \citep{Gouin_2021, Lau_2021, Cataldi_2023} have also been linked to its MAH. 

However, most of these works correspond to numerical studies, where the whole evolutionary history of these structures can be explicitly followed. While there are recipes for the direct estimation of mass accretion rates (MARs) on actual observational data (e.g., \citealp{Pizzardo_2021, Pizzardo_2022}; using the caustic technique under reasonable, simplifying assumptions, \citealp{deBoni_2016}; or \citealp{Soltis_2024}, who build a machine learning model based on multi-wavelength data), most often the assessment of the assembly state of galaxy clusters is done by looking at morphological and dynamical properties of either their radio, optical or X-ray emission, or their Sunyaev-Zeldovich signal \citep[for instance,][]{Rasia_2013, Lovisari_2017, Zenteno_2020, Campitello_2022, Biava_2024}.

At the same time, since up to three decades ago, simulations have aided in trying to understand which properties of a cluster or its DM halo are most affected by its MAH. Simple quantities that can be derived from the three-dimensional description only available in simulations include the offset between any pair of definitions of cluster centre \citep{Crone_1996, Maccio_2007, Cui_2016}, the ratio of kinetic to gravitational energies \citep{Shaw_2006, Neto_2007, Knebe_2008, Cui_2017}, or the fraction of the total halo mass in substructures (\citealp{Neto_2007}; or variants thereof, e.g. \citealp{Kimmig_2023}). More recently, other quantities have shown a great potential in constraining evolutionary properties of DM haloes, such as halo concentration \citep{Neto_2007, Wang_2020} or sparsity \citep{Balmes_2014, Corasaniti_2018}, just to mention a few. Also the thermodynamical state of the cluster core (in particular, cool-coredness, \citealp{Rasia_2015, Rossetti_2017}) and properties of the stellar component (e.g., fossilness, \citealp{Ragagnin_2019}, or fraction of mass in the brightest cluster galaxy and the intracluster light, \citealp{Kimmig_2025}), are profusely studied in relation to cluster assembly.

Even beyond the intrinsic interest on this topic because of its implications for observational studies (e.g., sample selection, constraining assembly histories, etc.), the search for proxies for the dynamical state or the assembly information has been targeting growing interest lately, as it has been shown that this notion of dynamical state is multifaceted (\citealp{Haggar_2024}; in the sense of being able to be split in several, more fundamental properties highlighting different distinct, uncorrelated features) and potentially time-dependent (\citealp{Valles-Perez_2023}, hereon \citetalias{Valles-Perez_2023}; implying that the optimal combination of different indicators that best predicts the presence of assembly episodes is not constant through cosmic time, but rather strongly evolving, instead), where additionally the different properties under study tend to be only mildly correlated \citep{Jeeson-Danie_2011, Skibba_2011}. 

In this context, the recent literature on the topic still lacks a comprehensive and systematic view on \emph{what part} of the MAH of galaxy groups and clusters is each of the myriad of indicators informing about, and in \emph{how far} does it constrain the MAH or quantities derived from it. \citet{Wong_2012} performed one of the first studies in this particular direction, using DM-only simulations to explore how the MAH of DM haloes can be decomposed, and what information can halo shape yield on this decomposition. Also, \citet{Giocoli_2012} described how halo concentration depends, in a non-trivial way, on several epochs of the assembly history of the halo (i.e., on a combination of the times at which the halo had built $4\%$ and $50\%$ of its present-day mass).
However, to our best knowledge, no study in the literature has systematically looked at the relation of DM halo morphological and dynamical parameters on instantaneous properties such as accretion rates, and neither accounted for the differences in ICM and DM quantities.

This is the first article of a series aimed at studying the effects of mass accretion onto simulated galaxy clusters and groups on their morphological, dynamical, thermodynamical, and observational properties. In this first paper, we have aimed to study in more detail what precise information on the full MAH and the associated MARs is yielded by a number of properties of the DM halo and the ICM when measured at $z=0$ (or at another, fix cosmic time), so as to constrain \emph{when}, \emph{what} and \emph{how much} are each of these quantities informing about. In particular, the specific aims of this work have been to establish \textit{(i)} what is the effect of selecting clusters based on their assembly state indicators on their ensemble MAHs; \textit{(ii)} when is mass accretion affecting the value of each parameter at $z=0$ the most; \textit{(iii)} besides timing, what additional information on the MAH is each parameter encoding; and \textit{(iv)} how consistent are these indicators when measured from the DM distribution and from the ICM.

The paper is organised as follows. In Sect. \ref{s:methods} we describe the simulation data, our set of assembly state indicators, and the statistical treatment. Subsequently, in Sect. \ref{s:results} we describe and discuss our results, in relation to the four points above. Finally, Sect. \ref{s:conclusions} contains a summary of the main conclusions together with additional discussion and comparisons with the literature. The two Appendices \ref{s:app.mtree} and \ref{s:app.PCA}, respectively, discuss in more detail our strategy for building the merger trees and its impact on our results, and more technical details about the principal component analysis of the assembly histories.

\section{Methods}
\label{s:methods}

The results in this paper stem from the analysis of a Lambda cold dark matter ($\Lambda$CDM) $N$-Body + hydrodynamical simulation, whose technical details are covered in Sect. \ref{s:methods.simulation}. The halo catalogues and merger trees, and the characterisation of their assembly state are subsequently described in Sects. \ref{s:methods.haloes} and \ref{s:methods.indicators}, respectively. In Sect. \ref{s:methods.MAHs} we cover our determination of the assembly histories and accretion rates. Finally, Sect. \ref{s:methods.statistics} deals with several aspects of the statistical analyses in this work.

\subsection{Simulation}
\label{s:methods.simulation}

We have performed and analysed a cosmological, $\Lambda$CDM simulation of a cubic domain $L = 100 \, h^{-1} \, \mathrm{cMpc}$ on each side. The assumed background cosmology is specified by a dimensionless Hubble constant $h \equiv H_0 / (100 \mathrm{\, km \, s^{-1} \, Mpc^{-1}}) = 0.678$, matter and baryonic density parameters $\Omega_m = 0.31$ and $\Omega_b = 0.048$, respectively, and null spatial curvature ($\Omega_\Lambda = 1 - \Omega_m$), overall compatible with the \citet{Planck_2020} results.

The initial conditions were generated from an unconstrained realisation of a primordial power spectrum with index $n_s=0.96$ and amplitude yielding $\sigma_8 = 0.82$, with a baryonic transfer function at $z=1000$ \citep{Eisenstein_1998}, further evolved until $z_\mathrm{ini} = 100$ under the \citet{Zeldovich_1970} approximation. From $z_\mathrm{ini}$ until $z=0$, the fully non-linear evolution is followed with the \texttt{MASCLET} code \citep{Quilis_2004}, which combines a Eulerian, high-resolution shock-capturing (HRSC) treatment of the equations of hydrodynamics and an $N$-Body approach for DM based on the particle-mesh method, both within an Adaptive Mesh Refinement (AMR) scheme allowing to achieve high dynamical ranges.

The simulation is run with a base grid with $N_x^3 = 256^3$ cells, and $n_\ell^\mathrm{ini}=3$ levels of refinement already present on the initial conditions (equivalent to an effective $2048^3$ resolution in dense regions). During the evolution, up to $n_\ell = 6$ refinement levels are dynamically generated, with criteria based on overdensity (standard, pseudo-Lagrangian AMR), converging flows and Jeans length. Overall, this grants a peak spatial resolution of $\Delta x_6 = 9 \, \mathrm{kpc}$ and a peak DM mass resolution of $M_\mathrm{DM,peak} = 1.5 \times 10^7 M_\odot$. Besides hydrodynamical and gravitational forces, only cooling is included, but no star formation or additional feedback mechanisms. Further details on this simulation are also covered by \citet{Valles-Perez_2024_NatAstro}.

Since this work is focused on large aperture radii (typically, $R_{200m}$; see definition below), the lack of feedback mechanisms, which typically impact the inner radii ($r\lesssim(0.3-1)R_{500c}$; e.g., \citealp{Duffy_2010, Planelles_2014, Planelles_2017}), is unlikely to have a large qualitative contribution to our results. Masses (and, thus, MAHs) at $R_{200m}$ are minimally affected by feedback (e.g., \citealp{Teyssier_2011}). Regarding cluster properties, such as those introduced below in Sec. \ref{s:methods.indicators}, while feedback could moderately impact their values, they are quantities integrated over a large ($\sim R_\mathrm{vir}$) aperture and should therefore remain relatively stable, and be monotonically related to the equivalent quantities in the absence of feedback (albeit with considerable scatter; see, for instance, \citealp{Corasaniti_2025} for the case of sparsity).

\subsection{Cluster catalogues and merger trees}
\label{s:methods.haloes}

Simulations are post-processed with the public halo finder \texttt{ASOHF}\footnote{\url{https://www.github.com/dvallesp/ASOHF}.} \citep{Planelles_2010, Knebe_2011, Valles-Perez_2022} to extract a catalogue of DM haloes at each snapshot. Haloes at different snapshots are connected using the ancillary \texttt{mtree.py} code of the \texttt{ASOHF} package, which determines all contributors to a halo in an (immediate or not) previous snapshot.

The catalogue of galaxy clusters and groups is then selected at $z=0$ as a mass-limited sample, containing all objects with virial mass\footnote{Here, $M_\mathrm{vir}$ is the virial mass using the prescription of \citet{Bryan_1998}, which corresponds to the mass in a spherical overdensity $\Delta_c = {18 \pi^2 + 39 \left[ \Omega_m(z) - 1\right] - 82 \left[ \Omega_m(z) - 1\right]^2 }$ with respect to the critical density $\rho_c = 3 H^2 / (8 \pi G)$, $G$ being the gravitational constant. At $z=0$, this corresponds to $\Delta_{\mathrm{m,vir}} \approxeq 330.7$.} $M_\mathrm{vir}^\mathrm{DM}(z=0) > 10^{13} M_\odot$, which are additionally sufficiently well-refined (at least up to $\ell = 4$, $\Delta x_4 = 36 \, \mathrm{kpc}$) and away from the domain boundaries. The main branch of the merger tree is traced back in time down to $z=4$ using a double-fold criterion looking the most gravitationally bound particle in the progenitor and descendant haloes (i.e., the most-bound particle in the progenitor is amongst the descendant particles and vice-versa; see section 2.6.2 in \citealp{Valles-Perez_2022}, as well as App. \ref{s:app.mtree} here, for more details and discussion).

The resulting catalogue comprises 31 galaxy clusters, whose $z=0$ virial mass exceeds $10^{14} M_\odot$, together with 358 groups ($10^{13} < M_\mathrm{vir}^\mathrm{DM} / M_\odot < 10^{14}$). Regarding the completeness of the merger tree, $90\%$ of the catalogue is traced back to at least ${z=2}$. The fraction starts to decline more rapidly at higher redshift, with $80\%$ at $z=3$ and $50\%$ at $z \gtrsim 4$. Unless where specified otherwise, the results are obtained by considering the complete sample of clusters and groups.

\subsection{Assembly state indicators}
\label{s:methods.indicators}

We aim to study how the MAHs of our clusters and groups depend on several parameters of the DM (or the ICM) at given, fixed cosmic times. Many of these indicators were already introduced by \citetalias{Valles-Perez_2023} in their study of the assembly state of DM haloes, and are only summarised here for completeness.

\begin{itemize}

\item \textbf{Centre offset ($\Delta_r$)}, defined as the distance between the DM density peak and the centre of mass in units of the virial radius, $\Delta_r \equiv |\vec{r}_\mathrm{peak} - \vec{r}_\mathrm{CM}|/R_\mathrm{vir}$ (e.g., \citealp{Crone_1996, Lacey_1996, Cui_2016}).

\item \textbf{Virial ratio ($\eta$)}, i.e. the quotient between twice the kinetic and the gravitational binding energies, $\eta \equiv 2 T / |W|$. While there are different choices as per whether to correct this quantity by a surface energy term \citep{Poole_2006, Shaw_2006, Power_2012}, as discussed in \citetalias{Valles-Perez_2023}, we do not consider this correction, since it tends to erase the correlation with merging activity and would make this quantity less useful as an indicator of \textit{assembly state}.

\item \textbf{Mean radial velocity ($\langle \tilde v_r\rangle$)} of the dark matter particles, computed as 

\begin{equation}
    \langle \tilde v_r\rangle = \frac{1}{V_\mathrm{circ,vir} }\frac{\big|\sum_\alpha m_\alpha \vec{v_\alpha} \cdot \vec{\hat u_r}\big|}{\sum_\alpha m_\alpha},
\end{equation}

\noindent where $m_\alpha$ and $\vec{v_\alpha} \cdot \vec{\hat u_r}$ are, respectively, the mass and the radial velocity (in the halo reference frame) of the $\alpha$-th particle, while $V_\mathrm{circ,vir} = \sqrt{G M_\mathrm{vir} / R_\mathrm{vir}}$ is the circular velocity at the virial radius.

\item \textbf{Sparsity ($s_{200c,500c}$)}, which is the quotient between\footnote{We adopt the usual notation for $M_{\Delta c[m]}$ being the mass in a sphere of radius $R_{\Delta c[m]}$ enclosing an overdensity $\Delta$ times the critical [background] density of a $\Lambda$CDM cosmology.}$M_{200c}$ and $M_{500c}$ and provides similar information to halo concentration, albeit in a non-parametric way \citep{Balmes_2014, Corasaniti_2018, Richardson_2022}.

\item \textbf{Ellipticity ($\epsilon$)}, computed from the reduced shape tensor,

\begin{equation}
    S_{ij} = \sum_\alpha \frac{m_\alpha r_{\alpha,i} r_{\alpha,j}}{r_\alpha^2},
\end{equation}

\noindent whose eigenvalues $\lambda_1 \geq \lambda_2 \geq \lambda_3$ are proportional to the semiaxes ($a \geq b \geq c$) squared, from which we define $\epsilon \equiv 1 - \frac{c}{a}$. This and other quantities relating to halo shape have been recently related to the assembly history of DM haloes \citep[e.g.][]{Chen_2019, Lau_2021}.

\item \textbf{Substructure fraction ($f_\mathrm{sub}$)}, i.e. the quotient between the total mass in substructures (each delimited by its Jacobi radius; see, e.g., \citealp{Binney_1987}) over the total, virial mass of the halo.

\item \textbf{Spin parameter ($\lambda_\mathrm{Bullock}$)}, which is a normalised measure of the DM angular momentum in units of $L_\mathrm{vir} = M_\mathrm{vir} R_\mathrm{vir} V_\mathrm{circ} = \sqrt{G M_\mathrm{vir}^3 R_\mathrm{vir}}$ \citep{Bullock_2001}.

\item \textbf{Combined relaxedness indicator ($\chi$)}. Inspired by the functional form used by \citet{Haggar_2020} and subsequent works, in \citetalias{Valles-Perez_2023} we introduced a summary measure of relaxedness that takes into account the first six parameters in this list in a redshift-dependent combination. Formally, it is defined as 

\begin{equation}
    \chi = \left[ \sum_{i=1}^6 w_i (z) \left( \frac{X_i}{X_i^\mathrm{thr}(z)}\right)^2\right]^{-1/2},
    \label{eq:combined_indicator}
\end{equation}

\noindent where $X_i$ represents each of the six aforementioned parameters and the weights $w_i(z)$ and thresholds $X_i^\mathrm{thr}(z)$ were calibrated as a function of redshift by \citetalias{Valles-Perez_2023} and \citet{Valles-Perez_2024_Thesis}.

In this definition, $\chi$ does not account for $\lambda_\mathrm{Bullock}$. While the methodology in \citetalias{Valles-Perez_2023} admits introducing as many indicators as desired, $\lambda_\mathrm{Bullock}$ was not considered in the calibration by \citet{Valles-Perez_2024_Thesis}. For the sake of conciseness, we choose to stick to the aforementioned calibration in this work. Since the parameters for $\chi$ are calibrated to correlate with merging activity, we do not expect the inclusion of additional indicators to substantially impact our results.

\end{itemize}

Each of these quantities (except for $\chi$) is intuitively correlated with the notion of a dynamically active halo, which is undergoing significant assembly activity. All quantities are computed from the three-dimensional description of the DM distribution and, unless otherwise stated (e.g., $s_{200c,500c}$), all bound particles within the virial radius are used in the computation. 

\subsection{Determination and stacking of the MAHs and MARs}
\label{s:methods.MAHs}

The MAHs of the DM component for individual clusters/groups are determined by summing the masses of all bound DM particles within the spherical overdensity radius (e.g.\footnote{We opt for $R_{200m}$ as the aperture to define our MAHs, instead of the $R_\mathrm{vir}$ described in Sec. \ref{s:methods.haloes} to define the cluster sample, since the latter would introduce an explicit pseudoevolution of the MAHs due to the time-dependent overdensity.}, $R_{200m}$) at each snapshot with redshift $z$, yielding $M_{200m}^\mathrm{DM}(z)$. For gas mass, $M_{200m}^\mathrm{gas}(z)$, we add the mass of all cells whose centre lies within a sphere of $R_{200m}$. Here, treating cells as pointwise particles for the purpose of mass integration implies quasi-Poissonian errors of $\sim 1\%$ or below for the largest mass objects (up to $\sim 3\%$ for the lowest mass ones), which can be safely disregarded given the typical mass variations during the considered redshift interval of around 2 orders of magnitude. Note that we do not perform any unbinding for gas for simplicity, as baryons have already been shock-heated at distances much larger than the apertures we use and their velocity distribution does not exhibit the same high-velocity tails as DM.

The MAHs from a subsample of clusters (in terms of, e.g., mass or any assembly state indicator) are stacked at each redshift by iteratively computing the biweight mean (i.e., robust mean; e.g. \citealp{Beers_1990, Wilcox_2012}) of the values of $\log_{10} M(z)/M_0$. Missing data (either clusters that are lost in a particular snapshot and recovered afterwards, or clusters where the main branch of the merger tree is interrupted) are ignored in this calculation. Through the paper, we refer to two measures of the scatter around these MAHs:

\begin{itemize}
    \item Scatter around the MAH, computed from the $1\sigma$ confidence interval in $\log_{10} M(z)/M_0$, where $\sigma$ is determined as the robust standard deviation \citep[or biweight midvariance;][]{Wilcox_2012}.
    \item Uncertainty in determining the mean MAH (i.e., standard errors), computed by taking the standard deviation of the distribution of mean values obtained by bootstrap resampling.
\end{itemize}

Finally, the instantaneous MARs,

\begin{equation}
    \Gamma_{200m} = \frac{\mathrm{d} \log M_{200m}}{\mathrm{d} \log a}
\end{equation}

\noindent are computed in a similar way to \citet{Valles-Perez_2020}, by resampling the $M(a)$ by cubic interpolation in logarithmic space to a vector of values of the scale factor $a$ which is logarithmically spaced with $\Delta \log a = 0.01 \, \mathrm{dex}$. Then, we apply a \citet{Savitzky_1964} cubic filter to compute the derivatives with window length of 15 points ($\Delta \log a = 0.15 \, \mathrm{dex}$), which roughly corresponds to a dynamical time. Thus, our $\Gamma_{200m}$ are instantaneous accretion rates smoothed over a dynamical time, to prevent the noisy behaviour of individual MAHs to contaminate the determination of the derivatives.

\subsection{Partial correlations and directions of maximal variation}
\label{s:methods.statistics}

In different sections of the results, we make extensive reference to different correlation coefficients amongst assembly state indicators and the MAHs or MARs. For clarity, we describe all of them in detail here.

\subsubsection{Linear and non-linear partial correlations}
\label{s:methods.statistics.partial}

In Sect.~\ref{s:results.MAHs_vs_dynstate}, we have aimed to establish the relation between MAHs or MARs and several properties of the cluster. Given that these quantities may have implicit dependencies with mass (see, e.g., Figs.~\ref{fig:MAH_mass} and~\ref{fig:corner} below), it is important to account for them so that we look at the correlations at fixed mass. While splitting objects in mass subsamples could alleviate this problem, it would come at the cost of greatly reducing the sample sizes and thus the statistical power. Instead, we account for these dependences by using \textit{partial correlation coefficients} \citep{Cramer_1946}, which account for the correlation between a pair of variables (namely, $x$ and $y$) while fixing one or more additional quantities ($z$). If the correlations are assumed to be linear, this comes down to finding the linear relations $x(z)$ and $y(z)$ and determining the Pearson correlation coefficients of the residuals $(\epsilon_x)_i = x_i - x(z_i)$ and $(\epsilon_y)_i = y_i - y(z_i)$. This can be developed to derive a general expression for the partial correlation between $x$ and $y$ fixing $z$,

\begin{equation}
    \rho_{xy|z} = \frac{\rho_{xy} - \rho_{xz} \rho_{yz}}{\sqrt{1 - \rho_{xz}^2}\sqrt{1 - \rho_{yz}^2}},
    \label{eq:partial_corr}
\end{equation}

\noindent where $\rho_{AB}$ is the regular Pearson correlation coefficient. Dropping the assumption of linearity, Pearson coefficients in Eq.~\ref{eq:partial_corr} can be substituted by Spearman rank correlation coefficients\footnote{Note that the Spearman rank correlation coefficients are just the Pearson coefficients between the rank variables. Hence, instead of testing for linearity, they indicate monotonicity.} if the relation is monotonic.

In other instances (e.g., Sect.~\ref{s:results.MAHs_biparametric.biparametric}) we have aimed at quantifying the ability of a third variable $z$ to break the scatter between a pair of variables $x$ and $y$ with an unknown, possibly non-monotonic underlying relation. Being non-monotonic, the approach above (i.e., computing $\rho_{yz|x}$) is no longer valid. In these cases, we perform a non-parametric fit to the $y=f(x)$ relation with smoothing cubic splines. In particular, we use the \texttt{FITPACK} implementation \citep{Dierckx_1995}, where the number of polynomials (the lower this number, the smoother the resulting interpolation is) is set as the lowest fulfilling the condition

\begin{equation}
    \sum_{i=1}^N \frac{1}{\sigma(y_i)^2} \left[ y_i - f(x_i) \right]^2 < N,
\end{equation}

\noindent $N$ being the number of data points, even though we have checked how slight variations on the smoothing criterion have negligible effect on our results. We can then compute the Spearman rank correlation between the residuals and the third variable, $z$. We shall refer to this quantity as the \textit{non-parametric residual correlation}.

\subsubsection{Directions of maximal variation in a scatter plot}
\label{s:methods.statistics.direction}

In Sect.~\ref{s:results.MAHs_biparametric.constraints}, we are interested in finding the direction, within a parametric space ($x$, $y$) in which a third variable $z$ varies, given some observations $\left\{(x_i, y_i, z_i)\right\}_{i=1}^N$. While there are several intuitively straightforward approaches for this aim (i.e., fitting a plane or using the partial correlation coefficients; e.g., \citealp{Baker_2022, Haggar_2024}), we found these solutions to be often unstable in our case due to the presence of outliers.

Instead, we devise a simple heuristic approach, where we consider the function $u_\alpha(x_i, y_i) = x_i \cos \alpha + y_i \sin \alpha$, where $\alpha \in [0, \pi[$, yielding the component of the vector $(x_i, y_i)$ along the unit vector $\mathbf{\hat n} = (\cos \alpha, \sin \alpha)$. Then, we identify the direction of maximal variation as

\begin{equation}
    \hat \alpha = \underset{\alpha}{\mathrm{argmax}} \; \rho_\mathrm{sp}(u_\alpha, z) 
    \label{eq:direction_maximal_variation}
\end{equation}

\noindent where $\rho_\mathrm{sp}$ represents the Spearman correlation coefficient. Additionally, the maximum value attained when $\alpha = \hat \alpha$ can be interpreted as a measure of the capability of $z$ in breaking the scatter in the $xy$ plane. Therefore, these results can be represented as a vector in the $xy$-space,
\begin{equation}
    \vec{\rho}(z) = \rho_\mathrm{sp}(u_{\hat \alpha}, z) (\cos \hat \alpha, \sin \hat \alpha),
    \label{eq:vector_variation}
\end{equation}

\noindent whose magnitude indicates the strength of the monotonic relation, and which points in the direction of variation of $z$ in the $xy$ space.

\subsubsection{Interpretation of the correlation coefficients}
\label{s:methods.statistics.interpretation}

The Pearson correlation coefficient of $x$ and $y$ has the natural interpretation of the fraction of scatter in $y$ that can be explained by the variable $x$ via a linear fit $y(x)$. By definition, the Spearman rank correlation has an equivalent interpretation on the rank variable, i.e. on the ranking of the data (the positions of values when sorted monotonically).

While univocally defined, these definitions can be slightly difficult to interpret. In cases where we are comparing two equivalent variables (i.e., in Sect.~\ref{s:results.gas_vs_DM}, where we study the relation between several indicators of dynamical state when determined from either gas or DM), it is also useful to introduce the \citet{Kendall_1938} rank correlation coefficient as an additional measure of monotonicity,

\begin{equation}
    \tau_K = \frac{n_\mathrm{conc} - n_\mathrm{disc}}{n_\mathrm{pairs}}
    \label{eq:kendall}
\end{equation}

\noindent where $n_\mathrm{pairs} = N(N-1)/2 = n_\mathrm{conc} + n_\mathrm{disc}$ is the number of unordered pairs amongst the $N$ data points, while $n_\mathrm{conc}$ ($n_\mathrm{disc}$) is the number of concordant (discordant) pairs, i.e. pairs where $(x_i - x_j)(y_i - y_j) > 0$ ($<0$). Purely monotonic relations will have $|\tau_K| = 1$, while completely unsorted datapoints will have $\tau_K=0$.

\section{Results}
\label{s:results}

Following the methodology described through Sect. \ref{s:methods}, in Sect. \ref{s:results.MAHs} we perform a general description and validation of the MAHs of our sample. Sect. \ref{s:results.MAHs_vs_dynstate} then explores the information yielded by each assembly state indicator on the MARs and MAHs, while Sect. \ref{s:results.MAHs_biparametric} explores how can the diversity of MAHs be summarised in a reduced number of parameters, and in how far do the assembly state indicators constrain these parameters. Finally, Sect. \ref{s:results.gas_vs_DM} deals with the possibility of relating the properties of the DM halo to the ones of the ICM.

\subsection{General description of the MAHs}
\label{s:results.MAHs}

\begin{figure}
    \centering
    \includegraphics[width=\linewidth]{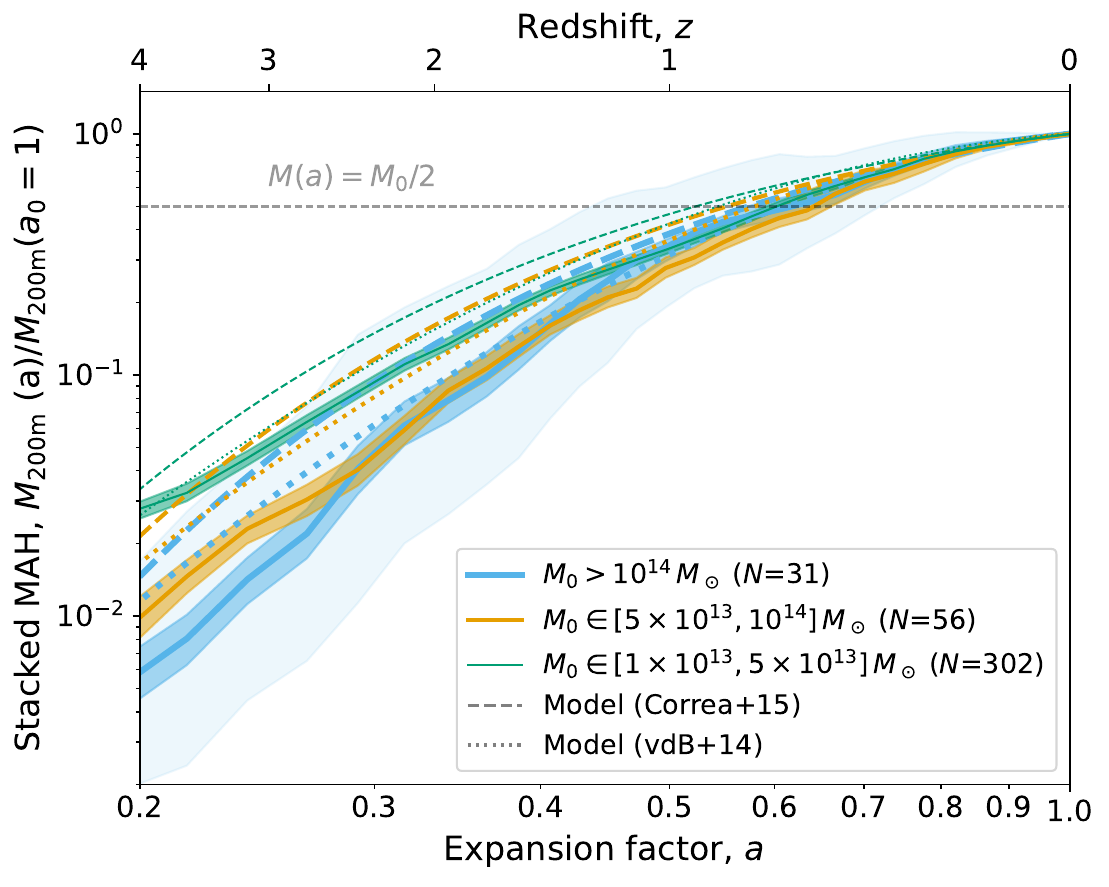}
    \caption{Stacked mass assembly histories (MAHs) in three mass subsamples, corresponding to galaxy clusters (light blue, thicker lines), high-mass groups (orange), and low-mass groups (turquoise, thinner lines). Dark shaded regions indicate the $1\sigma$ uncertainty around each mean MAH. Light shaded regions indicate the population standard deviation (just shown for the galaxy clusters subsample for clarity). Dashed (dotted) lines show the MAHs expected for each mass range according to the analytic (semi-analytic) model by \citet{Correa_2015} (\citealp{vandenBosch_2014}).}
    \label{fig:MAH_mass}
\end{figure}

As a first validation of our subsequent results, in Fig.~\ref{fig:MAH_mass} we show the total (dark + baryonic) stacked MAHs for three mass ranges at $z=0$, corresponding to clusters (light blue, thicker line; $M_0 > 10^{14} M_\odot$), high-mass groups (orange; $5 \times 10^{13} M_\odot < M_0 < 10^{14} M_\odot$) and low-mass groups (turquoise, thinner line; $10^{13} M_\odot < M_0 < 5 \times 10^{13} M_\odot$), determined according to the merger-tree and stacking procedures described in Sects.~\ref{s:methods.haloes} and~\ref{s:methods.MAHs}, respectively. In this figure, darker shaded regions indicate the standard error in the mean MAH, while broader, lighter shades enclose the standard deviation of the population of MAHs. 

Overall, our merger trees capture a stable trend of more massive haloes assembling later in cosmic time on statistical terms, i.e., $M(z) / M_0$ being a decreasing function of $M_0$ at fixed $z$, even though individual MAHs are very diverse and reach a $1\sigma$ scatter of $\sim 1 \, \mathrm{dex}$ at $z \simeq 4$. This is not reflected on typical measures of formation time (e.g., half-mass formation redshift, $z_{0.5}$; see the gray, dashed line intersecting with the ensemble MAHs) since differences only become significant over the uncertainty in the mean MAH at $z \gtrsim 1 \text{ and } 2$ for the low-mass (turquoise line) and high-mass (orange line) groups, respectively, when compared to the clusters sample (light blue line).

To show how our MAHs compare to other works, in the figure we also overplot the analytic predictions of \citet{Correa_2015} and the semi-analytic ones of \citet{vandenBosch_2014}, evaluated for each class at the median value of $M_0$. The comparison reveals noticeable differences with respect to these two models, where our MAHs lie up to a factor of 2 below the reference ones at high ($z \simeq 4$) redshift. As further discussed in App.~\ref{s:app.mtree}, these differences can be attributed to the strategy for selecting the main branch of the merger tree. Given the physical motivation of our strategy, based on a two-fold check using the most bound DM particles, we choose to stick to this choice, but warn the reader to be cautious when doing direct comparisons with the aforementioned references.

\subsection{Influence of assembly state parameters on the MAHs and MARs}
\label{s:results.MAHs_vs_dynstate}

In a first step of the analysis, we have aimed to understand the information provided by each of the assembly state indicators described in Sect.~\ref{s:methods.indicators}, by means of exploring the effect of selecting clusters by each indicator on their ensemble MAHs (computed and stacked as described in Sect.~\ref{s:methods.MAHs}).

\begin{figure*}
    \centering
    \includegraphics[width=\textwidth]{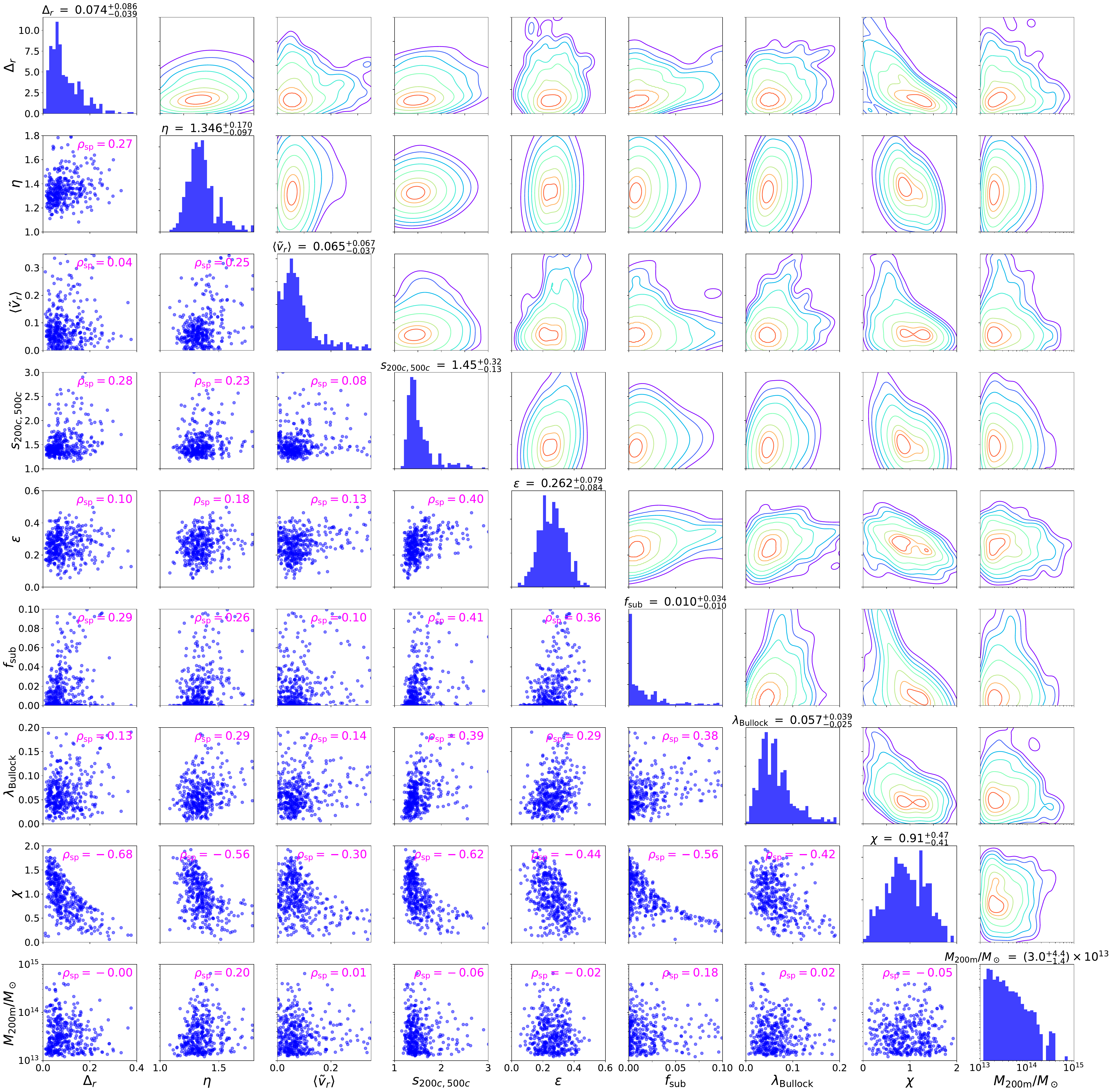}
    \caption{Corner plot showing the univariate and bivariate distribution of the assembly state indicators and mass at $R_{200m}$, all measured at $z=0$. Along the diagonal, each panel presents the distribution of one variable, with the text in the top indicating the median value and $(16,84)$ percentiles. The sub-diagonal terms are scatter plots of the pair of variables indicated by the respective row and column. The text in magenta within each off-diagonal panel indicates the Spearman rank correlation coefficient between the corresponding pair of variables. The panels above the diagonal represent the same information as contours, for better visualisation in the densest parts.}
    \label{fig:corner}
\end{figure*}

In Fig.~\ref{fig:corner}, we present a corner plot of the parameter space consisting of the eight individual indicators, together with cluster mass at $R_{200m}$, as a statistical summary of our basic dataset at $z=0$. Looking at the distribution of masses (bottom, right panel), our catalogue is naturally biased towards lower-mass systems, since it is a mass-limited sample. Looking at the bottom row, showing the bivariate distribution of all eight parameters with mass, no important mass dependences are found. The only significant ones ($p$-value $< 0.05$) are those with $\eta$ and $f_\mathrm{sub}$. While for the latter there could be numerical reasons (higher-mass haloes can contain a deeper hierarchy of substructures since they are sampled with more DM particles), the correlation of $\eta$ with mass may be purely physical and associated to the fact that more massive structures have assembled later on\footnote{We note that this weak correlation of $\eta$ with $M_{200m}$ vanishes if we account for the surface energy term in the computation of the virial ratio, as in \citet{Shaw_2006}. Nevertheless, the corrected ratio, $\eta'$, does not correlate with assembly state, as discussed above and in \citet{Valles-Perez_2023}.}. Interestingly, however, we do not find these mass dependencies in the rest of properties.

Amongst the assembly state parameters themselves\footnote{Excluding $\chi$, which is computed from the first six parameters in Sect.~\ref{s:methods.indicators} and, hence, is naturally highly correlated with each of them. In particular, for each value of a parameter, there is a maximum possible value for $\chi$ according to Eq. (\ref{eq:combined_indicator}).}, correlations are generally weak, perhaps with the exceptions of the pairs $f_\mathrm{sub}-\lambda_\mathrm{Bullock}$, $s_{200c,500c}-\lambda_\mathrm{Bullock}$, $f_\mathrm{sub}-s_{200c,500c}$, and $\epsilon-s_{200c,500c}$, which reach $\rho_s \simeq 0.4$. This highlights the complementarity of different probes of the assembly state which, even when recovered from the full three-dimensional description of simulations, tend to be highly scattered and loosely correlated \citepalias[see][for more discussion]{Valles-Perez_2023}. 

\subsubsection{Effects of the indicators at $z=0$ on the full MAHs}
\label{s:results.MAHs_vs_dynstate.MAHs}

\begin{figure*}
    \centering
    \includegraphics[width=0.39\textwidth]{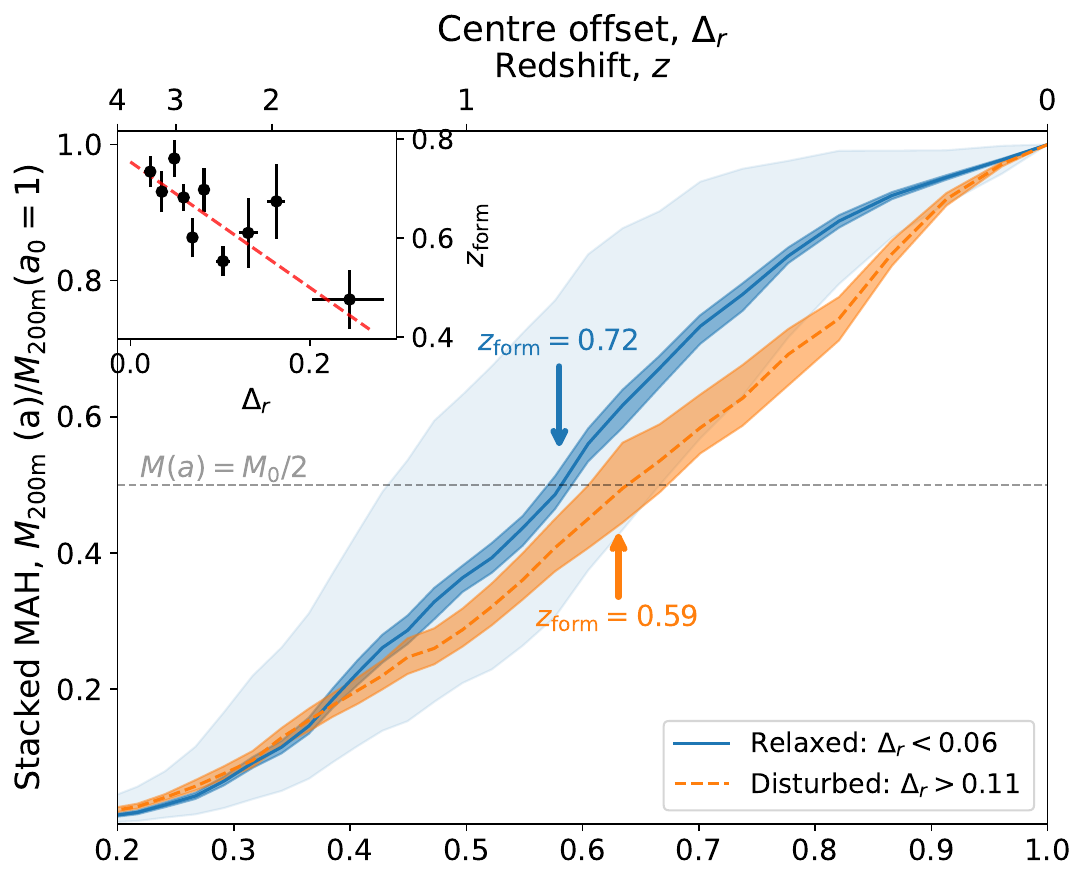}~
    \includegraphics[width=0.39\textwidth]{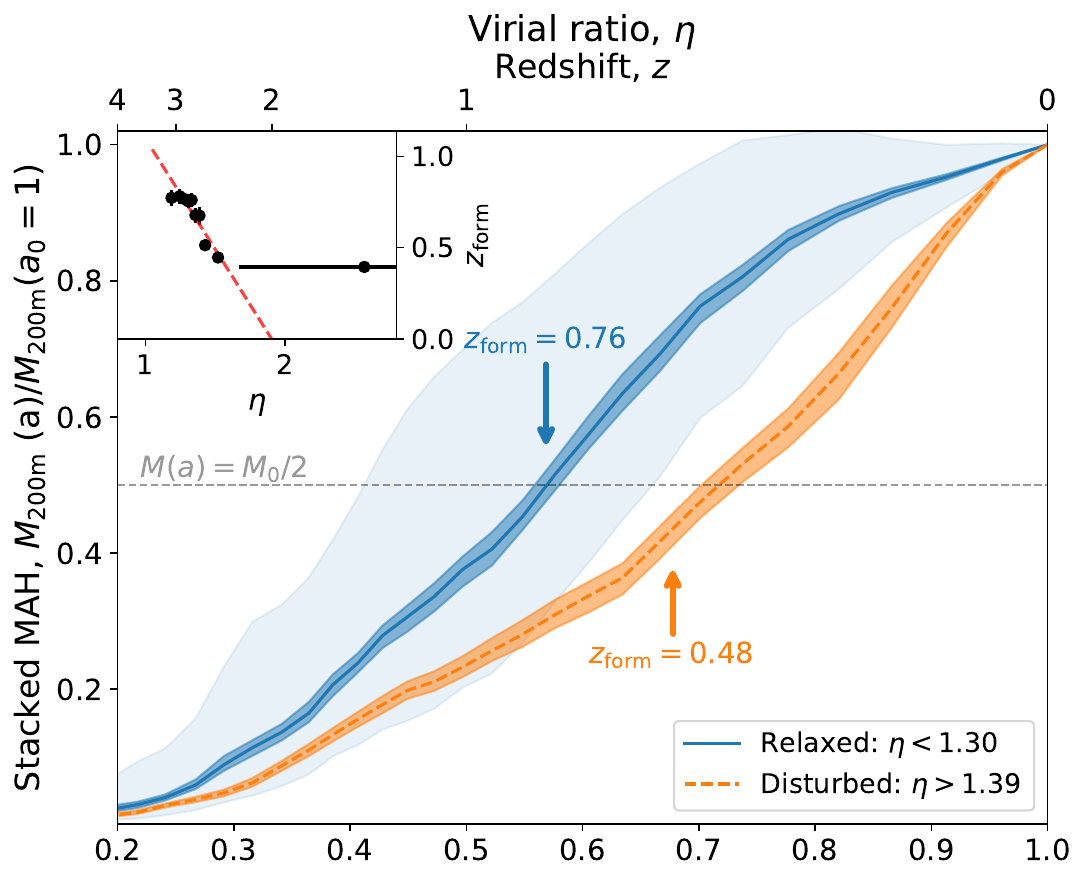}
    \includegraphics[width=0.39\textwidth]{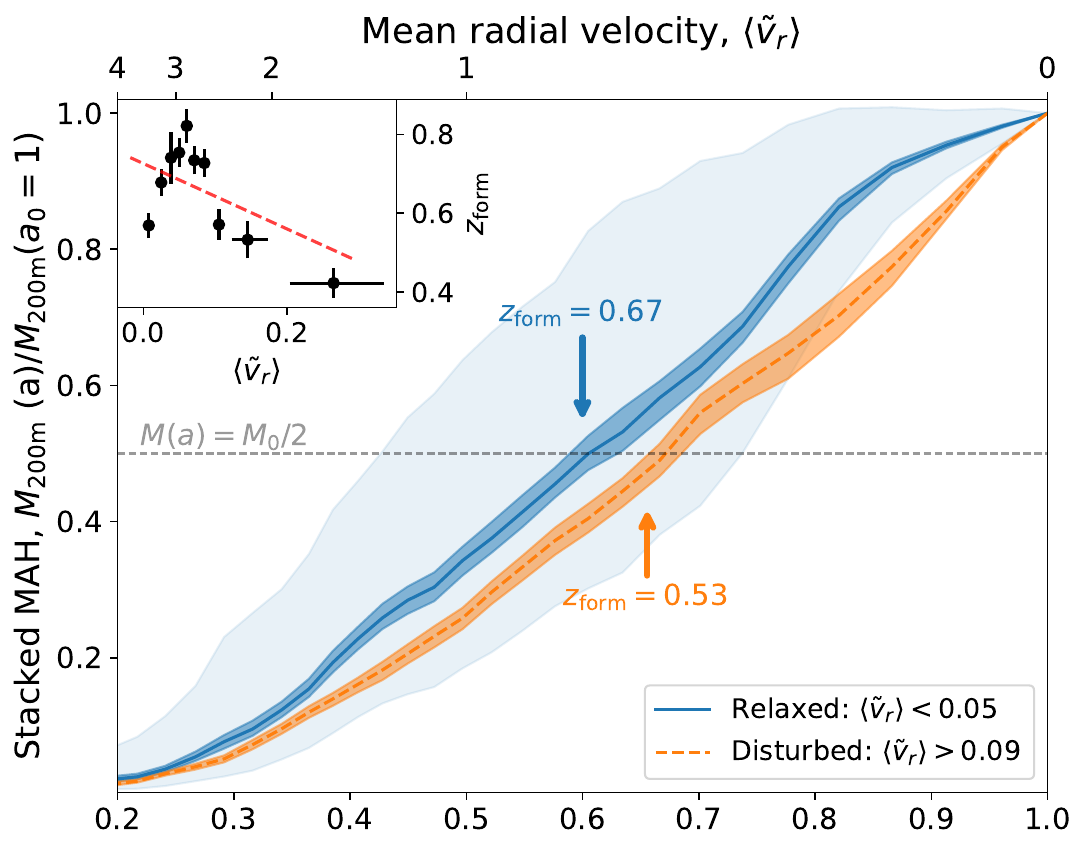}~
    \includegraphics[width=0.39\textwidth]{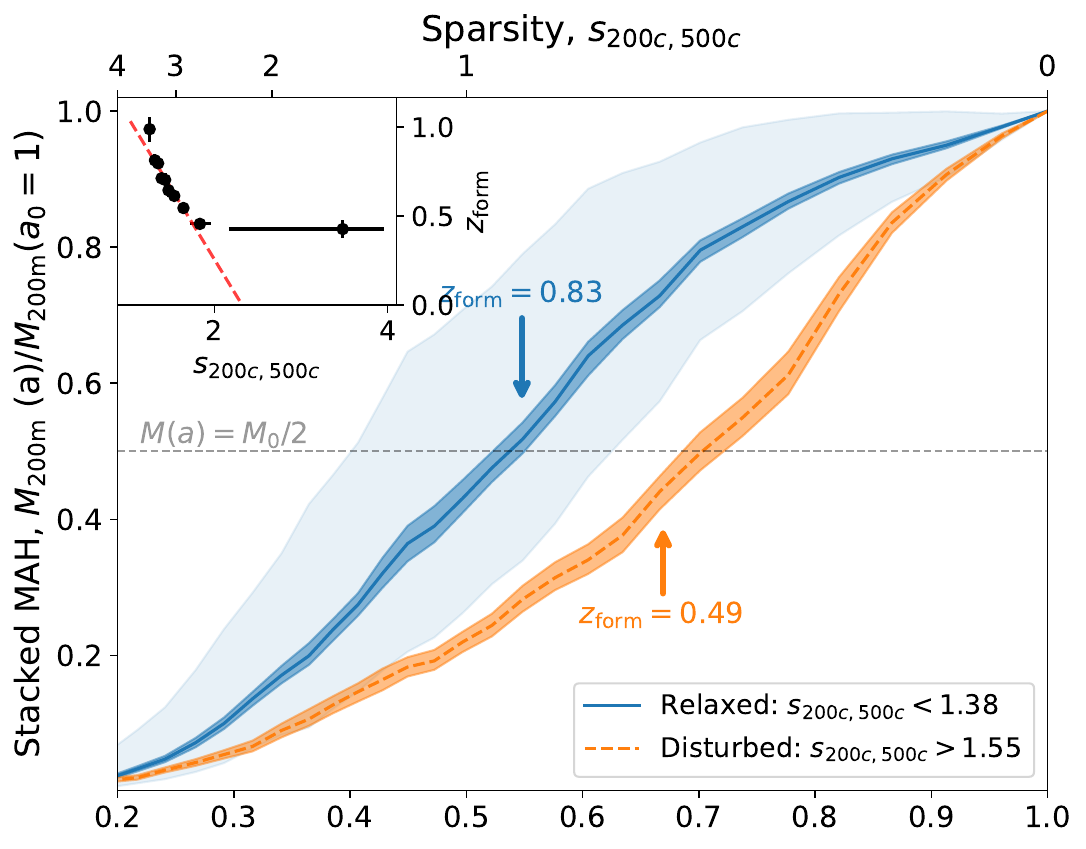}
    \includegraphics[width=0.39\textwidth]{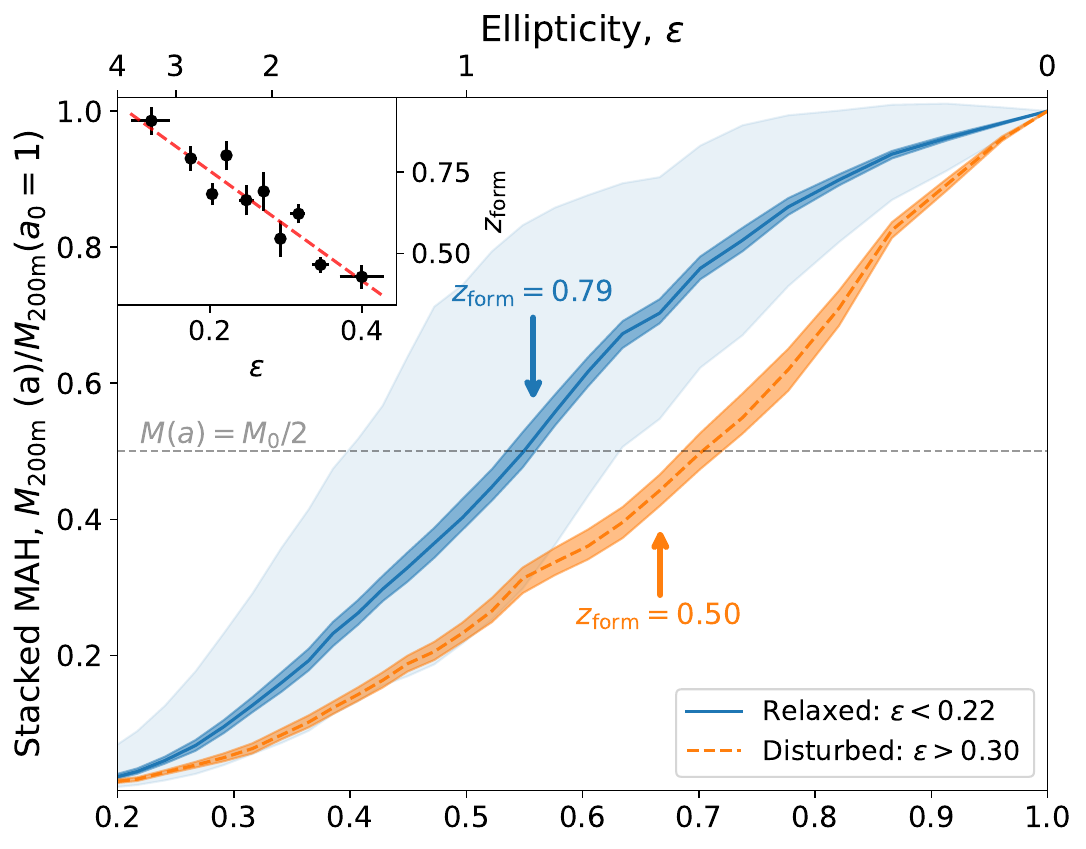}~
    \includegraphics[width=0.39\textwidth]{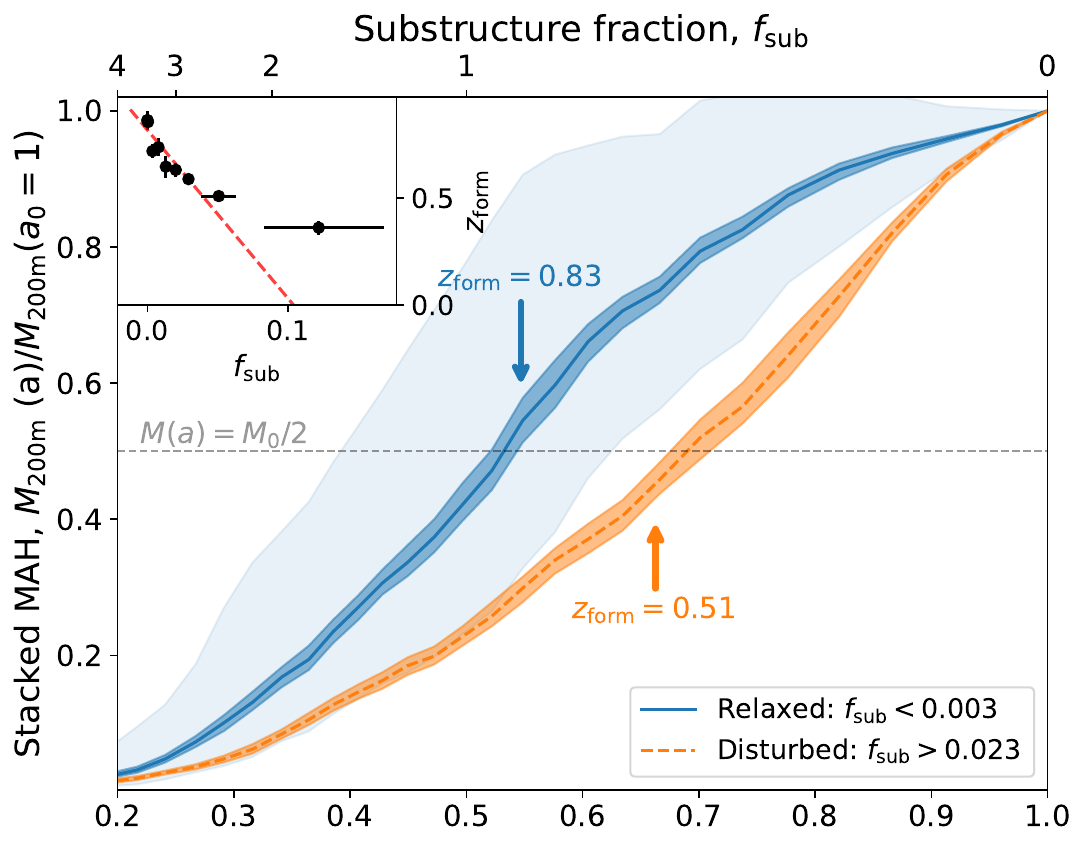}
    \includegraphics[width=0.39\textwidth]{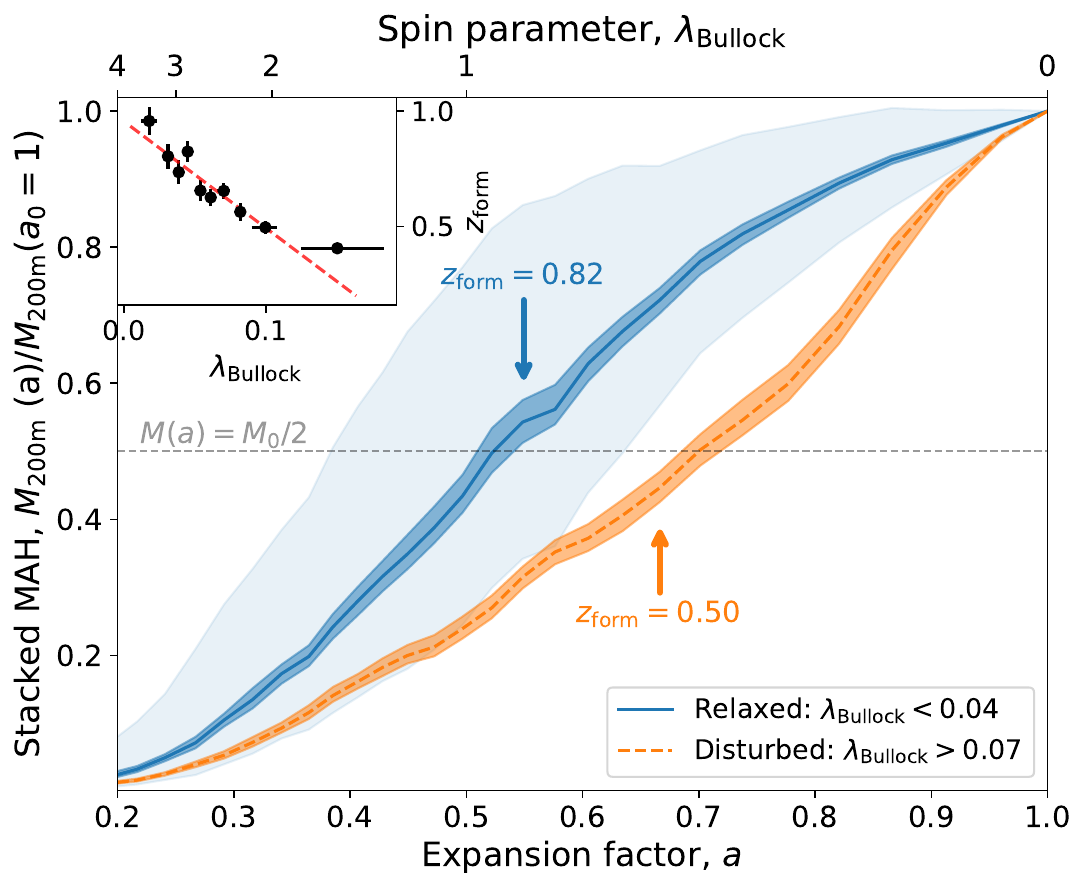}~
    \includegraphics[width=0.39\textwidth]{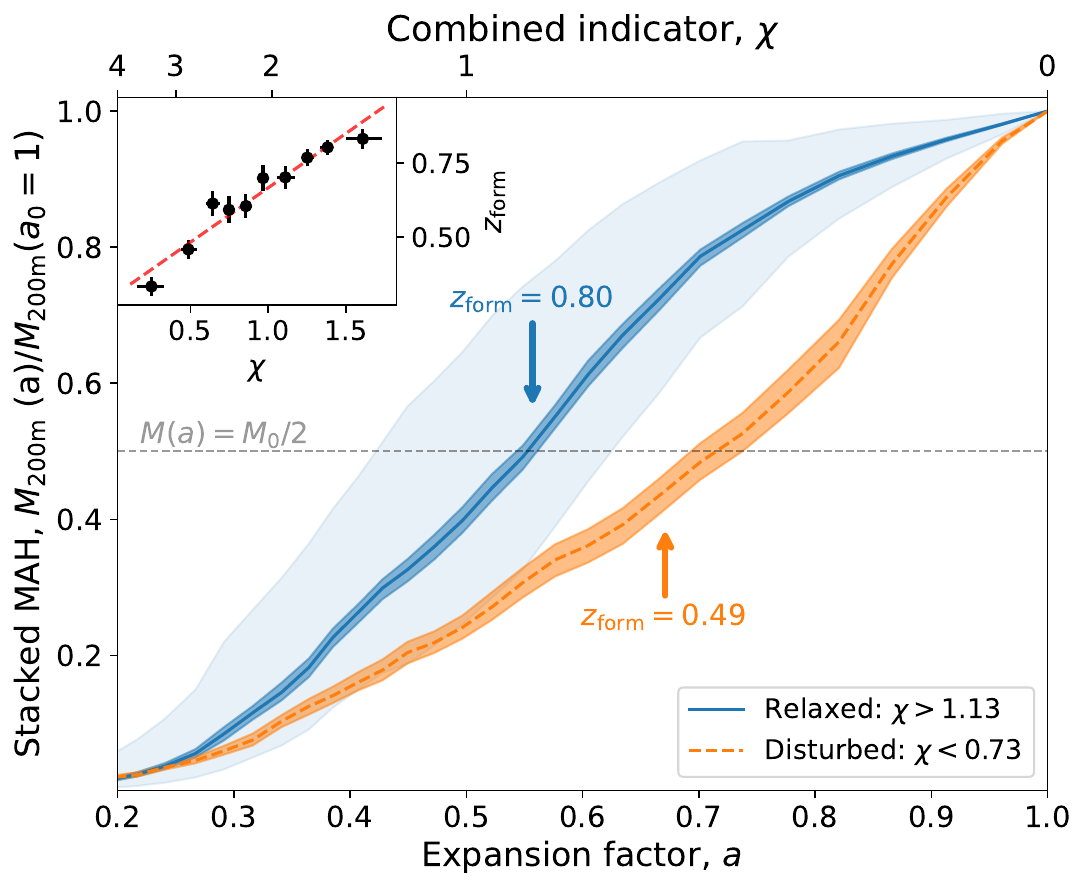}
    \caption{Effects of selecting haloes based on their assembly state indicators at $z=0$ on their full MAHs. Blue solid lines show the most relaxed third of the halo population at $z=0$ according to each parameter, specified in the subplot title. Orange dashed lines refer to the most disturbed third. Dark and light contours indicate uncertainty on the mean and population scatter, respectively. The insets indicate the dependence of $z_\mathrm{form}$ with the corresponding parameter, with a linear fit as the red, dashed line.}
    \label{fig:MAH_parameters}
\end{figure*}

In order to study what information about the full MAH is encoded by each indicator in Sect.~\ref{s:methods.indicators}, we have considered their ability in splitting the MAH curves. As a first indication of the relevance of the indicators, in the different panels of Fig.~\ref{fig:MAH_parameters} we present the stacked MAHs for the most-unrelaxed third (orange, dashed) and the most-relaxed third (blue, solid lines) according to each of the parameters in Sect.~\ref{s:methods.indicators} at $z=0$. Dark and light shaded regions indicate respectively the uncertainty in the class-averaged MAH and the population scatter around the MAH, as described in Sect.~\ref{s:methods.MAHs}. In these figures, we also indicate the formation redshift of each subsample, which we define from the formation time (the mass-weighted average of the accretion time of each mass element),

\begin{equation}
    t_\mathrm{form} = \frac{\int_{t(z_\mathrm{ini})}^{t_0} t \dot M \dd t}{\int_{t(z_\mathrm{ini})}^{t_0} \dot M \dd t},
    \label{eq:formation_time}
\end{equation}

\noindent from which we obtain $z_\mathrm{form} = z(t=t_\mathrm{form})$. Note that, while this definition roughly (but not exactly) agrees with the usual definition of $z_{0.5}$ as the time where $M(z_{0.5}) = M(z_0=0)/2$, it has the advantage of being defined from the whole MAH an hence is less sensitive to oscillations and, especially, to the coarse-graining of the simulation snapshots.

In general, even though the population scatter is large due to the huge diversity of assembly histories, the mean MAHs of the two extreme subsamples are always clearly separated in the natural direction (i.e., the disturbed subsample according to any parameter is more late-forming). However, there are clear variations of the amplitude and the location of such separation amongst indicators. As a rule of thumb for the following qualitative analysis, we shall indicate the separation to be relevant if the mean of one class falls outside the population scatter of the other.

For some indicators ($\Delta_r$ and $\langle \tilde{v_r} \rangle$) the separation of the MAHs is minimal and only focused at very low redshift ($a \gtrsim 0.8$), indicating that these parameters are only weakly related with the MAH for under a dynamical time. Wider effects on the MAHs are observed for the rest of indicators. 

It is interesting to compare these results to the ones shown in figure 4 of \citet{Haggar_2024}, where the authors report only noticeable differences in the mean MAHs when stacking them in groups of their \texttt{PC1} variable (mostly contributed by $f_\mathrm{sub}$ and $\Delta_r$), while no significant separation is obtained when using their \texttt{PC2} ($\eta$ and concentration; which, however, does encode the spread around the stacked MAH), \texttt{PC3} ($\lambda_\mathrm{Bullock}$ and cosmic web connectivity), or \texttt{PC4} (containing $\epsilon$). These findings, slightly at odds with our results that most of these parameters do effectively separate the MAHs, could possibly be due to the fact that their \texttt{PC} variables are linear combinations of many indicators, and their influence on the MAH could be lower than the individual variables if they contribute in opposite directions in the linear combination.

\begin{figure*}
    \sidecaption
    \includegraphics[width=12.5cm]{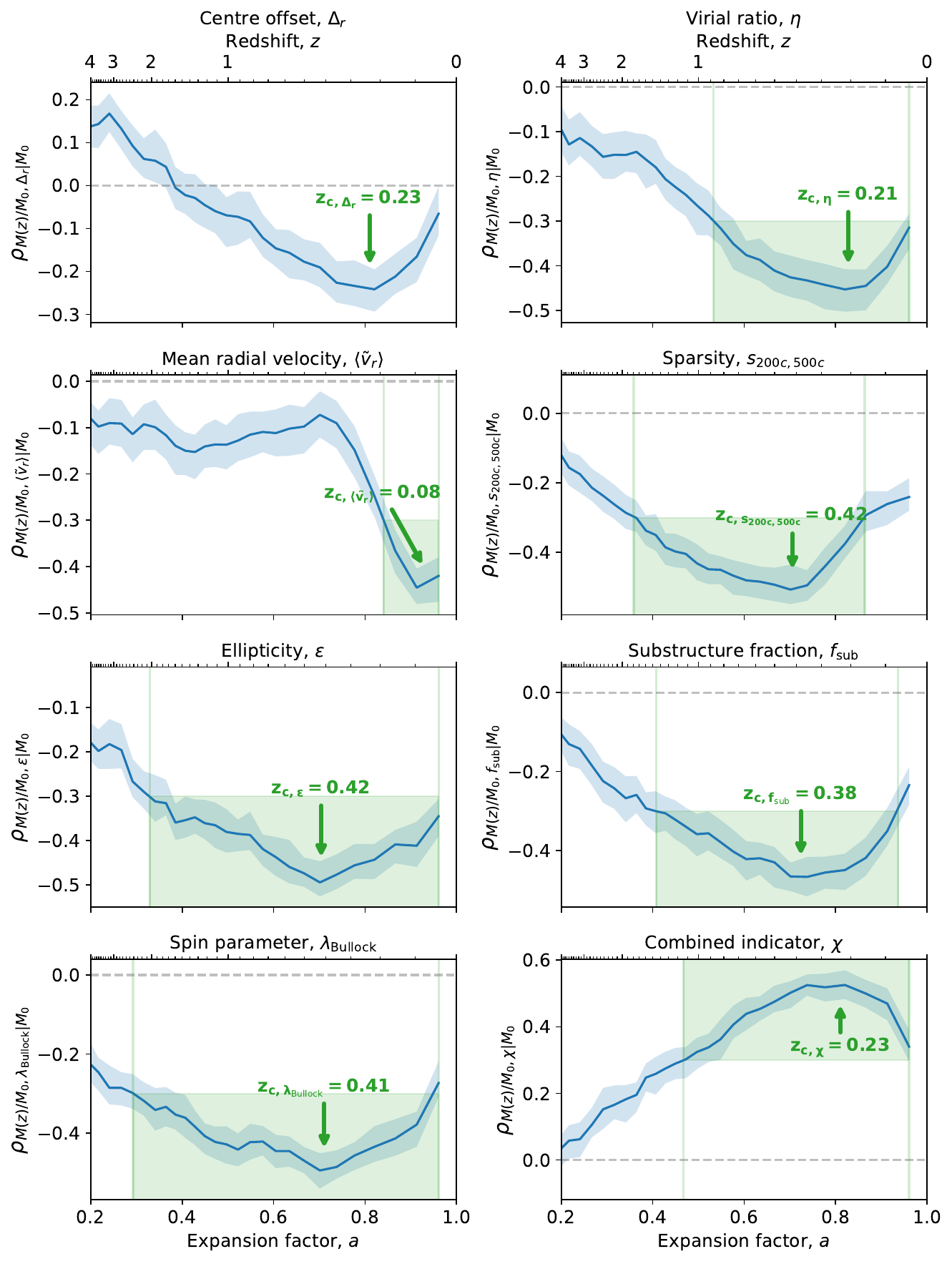}
    \caption{Constraining power of the $z_0=0$ assembly state indicators on the MAHs. Each panel shows the Spearman rank correlations between the MAH curves and a $z_0=0$ assembly state indicator. The absolute magnitude of the curve at each $z$ can be read as the fraction of the scatter on the ranking of $M(z)/M_0$ that can be explained by the corresponding parameter at $z_0 = 0$. The green regions indicate the redshift intervals where this correlation exceeds $0.3$ in magnitude, with the green, vertical lines delimiting the interval to ease the visualisation. The horizontal, grey, dashed line marks null correlation. The arrows indicate the redshift at which this correlation is maximal, $z_{c,X}$.}
    \label{fig:MAH_spearman}
\end{figure*}

However, little further quantitative information can be easily drawn from these panels. This is why we show, in Fig.~\ref{fig:MAH_spearman}, the Spearman rank-correlation coefficients between $M(z)/M_0$ and each indicator (measured at $z=0$), as a function of $z$. To rule out that any correlation is due to mass dependence on both $M(z)/M_0$ and the assembly state indicator, they are computed as partial correlation coefficients controlling for the dependence on $M_0$ (find a more in-depth description in Sect.~\ref{s:methods.statistics.partial}). While the figures show the complete information $\rho_{M(z)/M_0, X | M_0}$ for each indicator $X$, in order to simplify the discussion let us consider that selecting clusters based on $X$ at $z_0=0$ is relevant for $M(z)/M_0$ if $|\rho_{M(z)/M_0, X | M_0}| > 0.3$ at any given $z$. Though arbitrary\footnote{The results on the timing of the different indicators are by construction unchanged by this threshold (i.e., the curves and the green arrow in Fig. \ref{fig:MAH_spearman}). The choice of explaining $30\%$ of the scatter to be deemed as relevant is, by definition, arbitrary, and whether it is enough or not will inevitably depend on the application.}, this corresponds to identifying the interval of cosmic time where the value of the indicator at $z_0=0$ explains over $30\%$ of the scatter on the ranking of the values of $M(z)/M_0$. Here, it is easy to distinguish three groups of indicators:

\begin{itemize}
    \item \textit{Late-time} indicators, most linked to the MAH at $z \simeq 0.1$ and no further back than $z \simeq 0.2$ ($\simeq 2 \tau_\mathrm{dyn}/3$ back\footnote{$\tau_\mathrm{dyn}(z)= \left[ G \rho \right]^{-1/2} = \left[ G \rho_B(z) \Delta_\mathrm{vir}(z) \right]^{-1/2}$ is the dynamical timescale associated to the virial density of a collapsed DM halo, where $\rho_B(z) = \Omega_m(z) \rho_c(z)$ is the background matter density and $\Delta_\mathrm{vir}(z)$ is the virial overdensity given by \citet{Bryan_1998}.}): the mean radial velocity $\langle \tilde{v_r} \rangle$.
    \item \textit{Mid-time} indicators, being sensitive to $M(z)/M_0$ in the broad redshift interval $0 \lesssim z \lesssim 1$, with the correlation peaking at $z \simeq 0.2-0.25$ ($[0.6-0.8]\tau_\mathrm{dyn}$ ago): the virial ratio $\eta$, our combined indicator $\chi$, and, although at a much less significant level, the centre offset $\Delta_r$.
    \item \textit{Early-time} indicators, with their peak correlation with the MAH happening at $z \simeq 0.4$ (over a dynamical time ago), and being influenced by the past MAH up to $z \simeq 3$: sparsity $s_{200c,500c}$, ellipticity $\epsilon$, substructure fraction $f_\mathrm{sub}$ or the spin parameter $\lambda_\mathrm{Bullock}$.
\end{itemize}

It is important noting that a given parameter at $z=0$ correlating with the MAH at $z \simeq 3$ does not equate to the fact that mass accretion at $z\simeq 3$ is conditioning the value of the parameter at $z=0$. This is discussed in greater detail below, in Sect. \ref{s:results.MAHs_vs_dynstate.MARs}.

While, ideally, we could aim to explicitly parametrise the secondary dependence of the MAH on each assembly state indicator $M(a | X)$, this is unfeasible in the present study due to the limited statistics together with the fact that the $M(a)$ values already depend on $M_0$ and we are considering the whole mass range $10^{13} M_\odot \lesssim M_\mathrm{vir,0} \lesssim 10^{15} M_\odot$. 

Nevertheless, as a first step towards this goal, in the inset of each panel in Fig.~\ref{fig:MAH_parameters} we show how $z_\mathrm{form}$ varies continuously with the corresponding indicator, as obtained from stacking of the MAHs within each decile of the indicator. In this way, even in cases where the separation between the MAHs was not obvious, we find a clear trend for $z_\mathrm{form}$ (e.g. in the case of $\Delta_r$; cf. \citealp{Power_2012}, their figure 9).

These dependences can be to first order fitted to linear relations, so as to get the primary trends of formation redshift (and hence, of the MAH; see also Sect.~\ref{s:results.MAHs_biparametric}) with halo properties at $z=0$. These results, $z_\mathrm{form}(X)$, together with their $1\sigma$ uncertainties, $\sigma_{z_\mathrm{form}}(X)$, the validity region of each fit and their $R^2$ values are summarised in Table~\ref{tab:fits_zform}.

\begin{table*}
    \centering
    \caption{Fitted relations for determining the formation redshift $z_\mathrm{form}$ from each of the assembly state indicators (in each row) introduced in Sect.~\ref{s:methods.indicators}. The third column gives the $1\sigma$ uncertainty in the determination of $z_\mathrm{form}$ of an individual halo. The fourth column informs about the validity regions for each fit, while the fifth column gives the coefficient of determination $R^2$ as a measure of the goodness of the fit $z_\mathrm{form}(X)$.}
    \small
    \begin{tabular}{c|c|c|c|c}
Parameter & Fit & Scatter & Validity region & $R^2$ \\ \hline
Centre offset, $\Delta_r$ & $z_\mathrm{form}(\Delta_r) = 0.75-1.33\Delta_r$ & $\sigma_{z_\mathrm{form}}(\Delta_r) = 0.18+1.15\Delta_r$ & $\Delta_r \in$ [0, 0.25] & 0.55 \\
Virial ratio, $\eta$ & $z_\mathrm{form}(\eta) = 1.07-1.17(\eta-1)$ & $\sigma_{z_\mathrm{form}}(\eta) = 0.23+0.10(\eta-1)$ & $\eta \in$ [1.1, 1.6] & 0.84 \\
Mean radial velocity, $\langle \tilde v_r \rangle$ & $z_\mathrm{form}(\langle \tilde v_r \rangle) = 0.73-0.89\langle \tilde v_r \rangle$ & $\sigma_{z_\mathrm{form}}(\langle \tilde v_r \rangle) = 0.24+0.53\langle \tilde v_r \rangle$ & $\langle \tilde v_r \rangle \in$ [0, 0.25] & 0.04 \\
Sparsity, $s_{200c}^{500c}$ & $z_\mathrm{form}(s_{200c}^{500c}) = 1.06-0.82(s_{200c}^{500c}-1)$ & $\sigma_{z_\mathrm{form}}(s_{200c}^{500c}) = 0.27-0.07 (s_{200c}^{500c}-1)$ & $s_{200c}^{500c} \in$ [1, 2] & 0.91 \\
Ellipticity, $\epsilon$ & $z_\mathrm{form}(\epsilon) = 1.10-1.71\epsilon$ & $\sigma_{z_\mathrm{form}}(\epsilon) = 0.38-0.47\epsilon$ & $\epsilon \in$ [0, 0.4] & 0.87 \\
Substructure fraction, $f_{\mathrm{sub}}$ & $z_\mathrm{form}(f_{\mathrm{sub}}) = 0.83-8.35f_{\mathrm{sub}}$ & $\sigma_{z_\mathrm{form}}(f_{\mathrm{sub}}) = 0.34-4.60f_{\mathrm{sub}}$ & $f_{\mathrm{sub}} \in$ [0, 0.08] & 0.89 \\
Spin parameter, $\lambda_{\mathrm{Bullock}}$ & $z_\mathrm{form}(\lambda_{\mathrm{Bullock}}) = 0.96-4.63\lambda_{\mathrm{Bullock}}$ & $\sigma_{z_\mathrm{form}}(\lambda_{\mathrm{Bullock}}) = 0.33-1.51\lambda_{\mathrm{Bullock}}$ & $\lambda_{\mathrm{Bullock}} \in$ [0, 0.1] & 0.86 \\
Combined indicator, $\chi$ & $z_\mathrm{form}(\chi) = 0.30+0.37\chi$ & $\sigma_{z_\mathrm{form}}(\chi) = 0.42-0.17\chi$ & $\chi \in$ [0, 1.5] & 0.95 \\
    \end{tabular}
    \label{tab:fits_zform}
\end{table*}

\subsubsection{What do indicators at a given $z$ tell about the MARs?}
\label{s:results.MAHs_vs_dynstate.MARs}

\begin{figure*}
    \sidecaption
    \includegraphics[width=12.5cm]{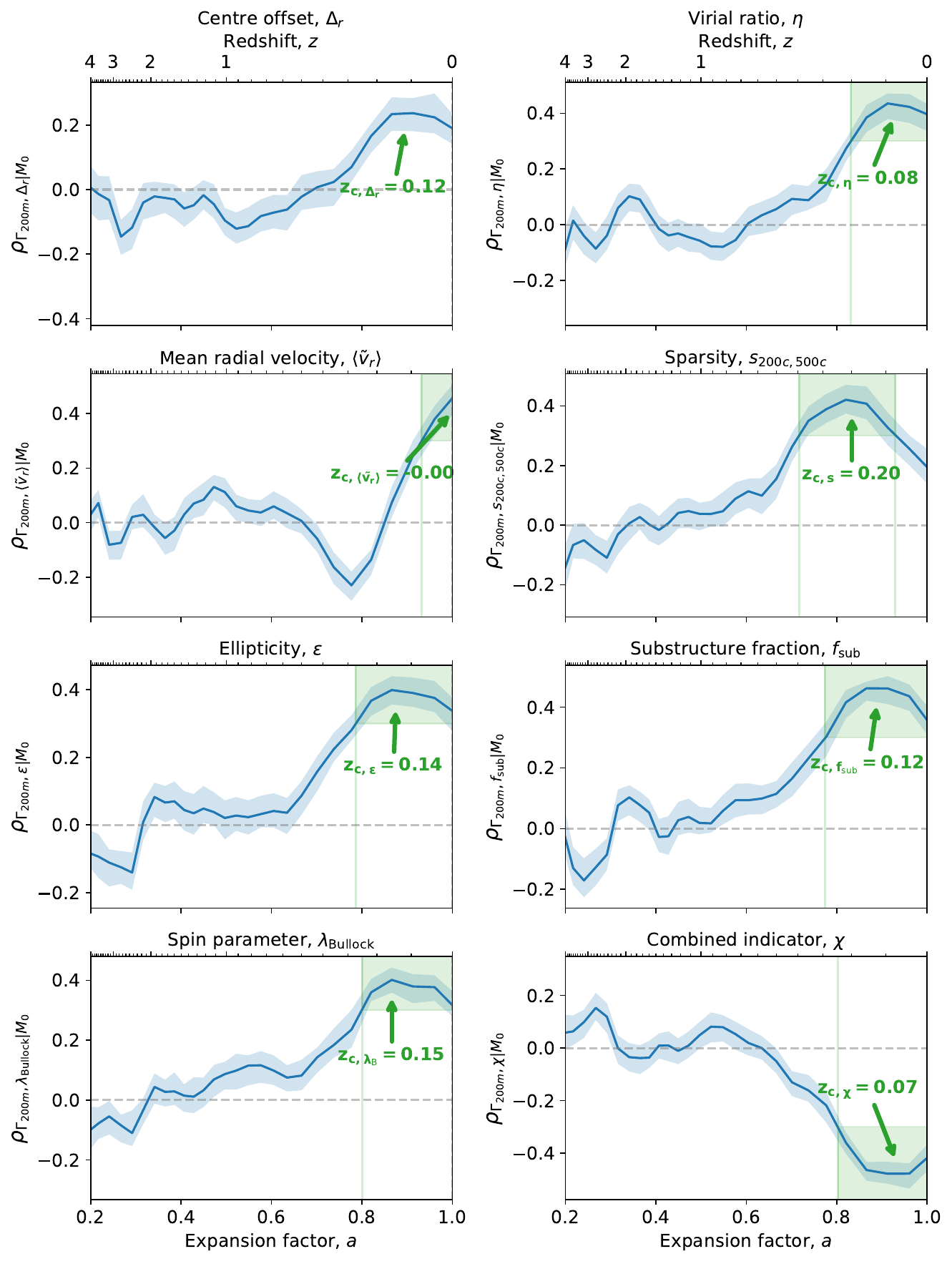}
    \caption{Impact of accretion rates on each of the $z_0=0$ assembly state indicators. Each panel shows the Spearman rank correlations between the instantaneous MARs and an assembly state indicator at $z_0 = 0$. All plot elements are equivalent to Fig.~\ref{fig:MAH_spearman}.}
    \label{fig:MARs_vs_dynstate_z0}
\end{figure*}

\begin{figure*}
    \sidecaption
    \includegraphics[width=12.5cm]{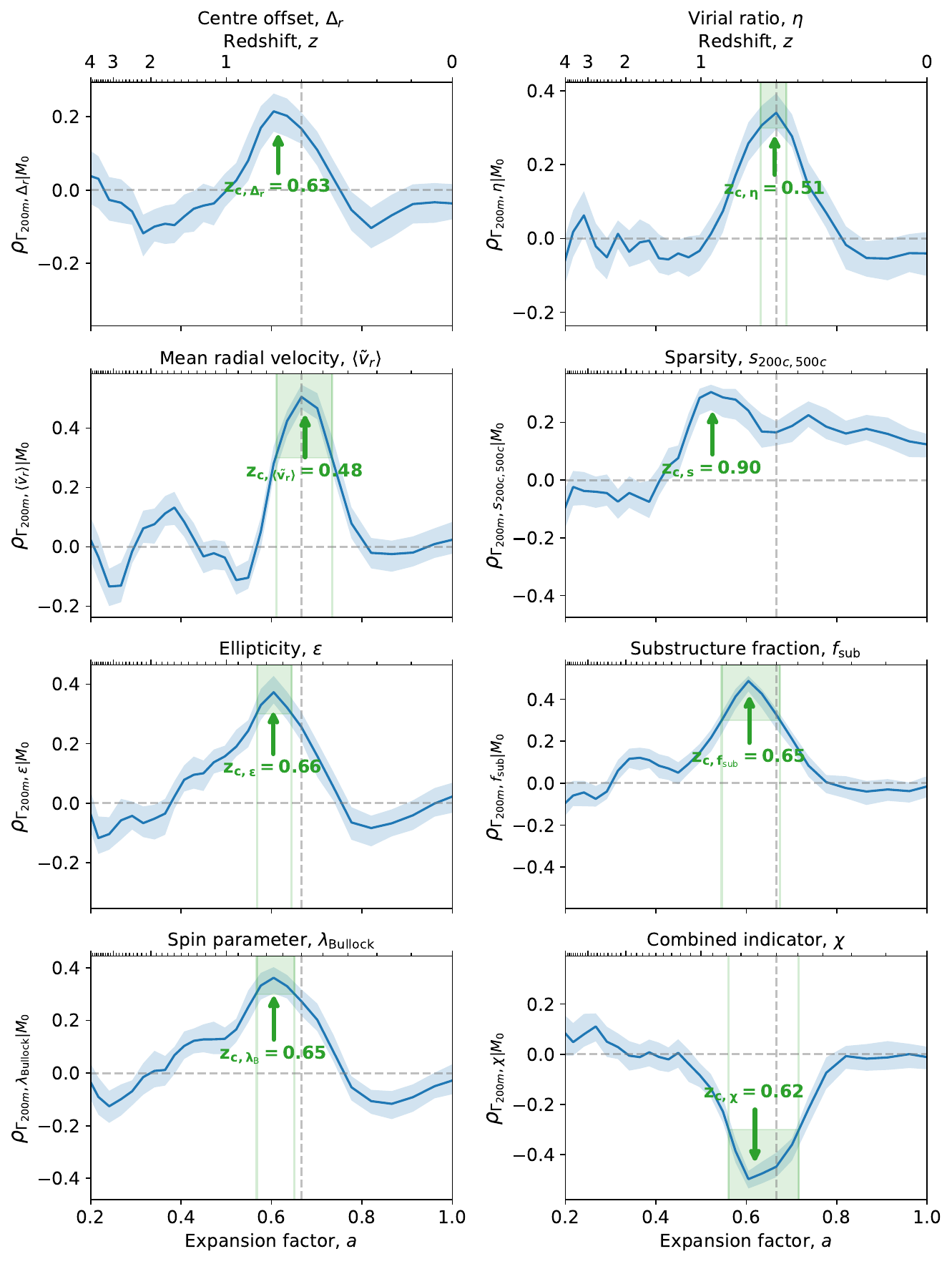}
    \caption{Impact of accretion rates on each of the $z=0.5$ (indicated as a vertical, grey, dashed line) assembly state indicators. All plot elements are equivalent to Fig.~\ref{fig:MARs_vs_dynstate_z0}.}
    \label{fig:MARs_vs_dynstate_z0.5}
\end{figure*}

While the previous study has focused on characterising the effect on the $M(z)$ curves of selecting clusters and groups based on their assembly state as inferred from different parameters, this is not to be confused by the characteristic epoch of accretion the parameter responds to. That is to say, the fact we have detected significant correlation between, e.g., $\lambda_\mathrm{Bullock}$ and $M(z)/M_0$ back to $z \simeq 2.5$ is not equivalent to saying that $\lambda_\mathrm{Bullock}$ at $z_0=0$ is related to the mass accreted at $z=2.5$. This is mostly due to the fact that the MAH curves are strongly self-correlated, in the sense that late-assembling clusters will display stronger slopes at low redshifts, and consequently lower values at high redshift. To complement this view, in this section we precisely explore the correlations between accretion rates $\Gamma_{200m}$ (defined in Sect.~\ref{s:methods.MAHs}) and single assembly state indicators at $z=0$ (Fig.~\ref{fig:MARs_vs_dynstate_z0}) and at $z=0.5$ (Fig.~\ref{fig:MARs_vs_dynstate_z0.5}), providing a more insightful picture so as to assess \textit{when} each indicator is informative about.

Focusing on the results for the parameters at $z=0$, in line with previous results in Fig.~\ref{fig:MAH_spearman}, $\Delta_r$ is incapable of providing sufficient insight on $\Gamma_{200m}$ (according to our arbitrary threshold of $0.3$ on the Spearman rank correlation magnitude). Also in this case there is a separation between indicators of earlier and later accretion: 

\begin{itemize}
    \item In particular, the mean radial velocity $\langle \tilde v_r \rangle$ appears to be the one most focused on present, instantaneous accretion, with $\rho_{\Gamma_{200m},\langle \tilde v_r \rangle | M_0}$ peaking at $z \simeq 0$.
    \item Like in the analysis of the MAHs, the virial ratio $\eta$ and the combined indicator $\chi$ are more mid-time indicators of assembly, being sensitive to accretion rates in $z \in [0, 0.2]$, peaking around $(0.2-0.3) \tau_\mathrm{dyn}$ ago.
    \item Substructure fraction $f_\mathrm{sub}$, ellipticity $\epsilon$ and spin parameter $\lambda_\mathrm{Bullock}$ are earlier indicators, being strongly correlated with accretion rates up to $z \simeq 0.3$ and with their influence being maximal around $\sim \tau_\mathrm{dyn}/2$ from the moment the indicators are determined.
    \item Lastly, sparsity stands as the earliest indicator, whose value at ${z_0=0}$ is strongly correlated with the MAR up to ${z = 0.4}$ (over a dynamical time ago), and especially at $z \simeq 0.2$ ($\sim~2\tau_\mathrm{dyn}/3$ ago)
\end{itemize} 

Among the indicators, the combined parameter $\chi$ (\citealp{Valles-Perez_2023}, \citealp{Valles-Perez_2024_Thesis}) provides the tightest constraints on the MAR (Spearman correlation magnitude $0.48$, implying that this quantity measured at $z=0$ can account for half of the scatter in the values of $\Gamma_{200m}$ at $z \simeq 0.07$), closely followed by the substructure fraction ($0.46$), although these refer to different epochs as reviewed above. 

In an analogous way, in Fig.~\ref{fig:MARs_vs_dynstate_z0.5} we analyse which accretion epochs are driving the values of these indicators at $z=0.5$. This has the additional interest of being able to assess the correlation between the indicators and, not only past accretion rate, but also future ones. At this earlier time, we can qualitatively discern at least two groups of variables:

\begin{itemize}
    \item Those responding to the accretion rates in an interval centred on $z=0.5$ (the instant when the parameters are determined), i.e., the indicators of \textit{ongoing} accretion. These include the virial ratio ($\eta$; roughly in $z \in [0.45, 0.6]$), the mean radial velocity ($\langle \tilde v_r \rangle$; $z \in [0.35, 0.65]$) and our combined indicator ($\chi$; for $z \in [0.4, 0.8]$). 
    
    The two latter ones show the strongest correlations ($\rho_{\Gamma_{200m},\langle \tilde v_r \rangle | M_0} = 0.49$ and ${\rho_{\Gamma_{200m},\chi| M_0} = 0.48}$, at their respective peaks).
    \item Those sensitive to past accretion rates, with their correlation peaking around $z \simeq 0.65$, i.e. around $\sim 0.4 \tau_\mathrm{dyn}$ before they are measured. This group consists on halo ellipticity ($\epsilon$; $z \in [0.55, 0.75]$), substructure fraction ($f_\mathrm{sub}$; $z \in [0.5, 0.85]$) and spin ($\lambda_\mathrm{Bullock}$; $z \in [0.55, 0.75]$).
\end{itemize}

Interestingly, even though it does not reach our threshold for significance at this higher redshift, sparsity presents its maximum correlation with the mass increase at $z \simeq 0.9$, corresponding to a full dynamical time ahead, also in line with the results at $z = 0$ presented above. This may be naturally explained from the fact that this quantity is sensitive to the density profile in between $R_{200c}$ and $R_{500c}$. Especially the latter, being a rather central region ($R_{500c} \sim [0.35-0.5] R_{200m}$), is affected with considerable delay by smooth accretion and mergers (unless these proceed with low impact parameter), whereas the other properties can start to be affected as soon as the infalling matter crosses the outer boundary of the halo.

Though subtle, it is important to study these relations at different redshifts through the assembly history of groups and clusters, since, as shown by \citetalias{Valles-Perez_2023}, the constraining power of these parameters on the presence/absence of assembly episodes varies strongly through cosmic history. While it would be interesting to pursue these analyses at higher redshift, in the present work we are limited by several factors, including the sparse saving of snapshots (ideally, we need around $6-10$ snapshots per dynamical time to be able to appreciate timing differences across indicators), or the increasing uncertainty of the determination of merger tree's main branches (which becomes less straightforward as we approach the protocluster regime beyond $z \sim 1.5-2$).

\subsection{Constraints on the MAHs from individual indicators at $z=0$}
\label{s:results.MAHs_biparametric}

Given the complexity and diversity of MAHs, we deal in Sect. \ref{s:results.MAHs_biparametric.biparametric} with the possibility of summarising this diversity with two parameters. Subsequently, in Sect. \ref{s:results.MAHs_biparametric.constraints}, we study how our assembly state indicators provide constraints in this biparametric space. 

All the results in this section correspond exclusively to the subsample of clusters that can be tracked back to $z_\mathrm{start}=4$. Such a selection may induce a slight bias for the MAHs, as it has been shown that major mergers pose an important challenge for halo finders and merger trees, particularly in recovering halo masses \citep{Behroozi_2015}. Nevertheless, we have explicitly verified that the discrepancy, at high redshift, between the MAH stacked over the whole cluster sample, and only those tracked back to $z > z_\mathrm{start}$, does not exceed $\sim 30\%$. Given that MAHs typically span over two orders of magnitude in our sample (see Fig. \ref{fig:MAH_mass}), this level of discrepancy is minor, and the approach remains robust.

\subsubsection{Biparametric characterisation of the MAHs}
\label{s:results.MAHs_biparametric.biparametric}

To reduce the dimensionality of the problem, we aim to capture as much as possible from the MAHs (a continuous function) with a minimal set of parameters. 
This was done, from the data-driven perspective, by e.g. \citet{Wong_2012}, who performed a principal component analysis (PCA; \citealp{Pearson_1901, Hotelling_1936}) on their assembly histories to extract and interpret the two most relevant contributions. Here, conversely, we have aimed to define two properties of the MAH with closed mathematical forms, so as to then explore to what extent do they capture the diversity of MAHs.

The first of such properties is, naturally, the formation redshift, $z_\mathrm{form}$, as defined by Eq. (\ref{eq:formation_time}). This quantity (or alternative definitions, such as $z_{0.5}$) has been widely used in the literature.

As a second parameter aimed at explaining the scatter in assembly histories at fixed formation time, we introduce a dimensionless quantity, $\alpha_\mathrm{MAH}$, measuring the episodicity of the assembly (i.e., whether it occurs suddenly in one major event, or more smoothly throughout the whole evolutionary history). Formally, we define it as

\begin{equation}
    \alpha_\mathrm{MAH} = \frac{t_{0.75} - t_{0.25}}{\tau_\mathrm{dyn}(z_\mathrm{form})},
\end{equation}

\noindent where $t_\beta$ is the latest time when $M(t_\beta) = \beta M_0$ ($0 < \beta \leq 1$). In this way, high values of $\alpha_\mathrm{MAH}$ indicate smooth assembly, and low values indicate bursty assembly around the formation time of the halo.\footnote{We have tested either normalising $t_{0.75} - t_{0.25}$ by $\tau_\mathrm{dyn}(z_\mathrm{form})$ and by $\tau_\mathrm{dyn}(z=0)$, finding that in the latter case there is a small but significant ($\rho_\mathrm{sp} \sim -0.3$) anticorrelation of the resulting parameter with $z_\mathrm{form}$. Hence, the former appears to be a more convenient choice to represent the MAHs.

It is also worth to mention that, while for $z_\mathrm{form}$ we chose not to use quantile-based measurements ($z_{0.5}$), this is not as problematic for $\alpha_\mathrm{MAH}$ since the noisy character associated to the coarse-graining of the snapshots is heavily suppressed by the fact that $t_{0.75}-t_{0.25}$ typically corresponds to $5-10$ snapshots.
}

\begin{figure*}
    \begin{minipage}[t]{0.7\textwidth}
        \centering
        \includegraphics[width=6cm]{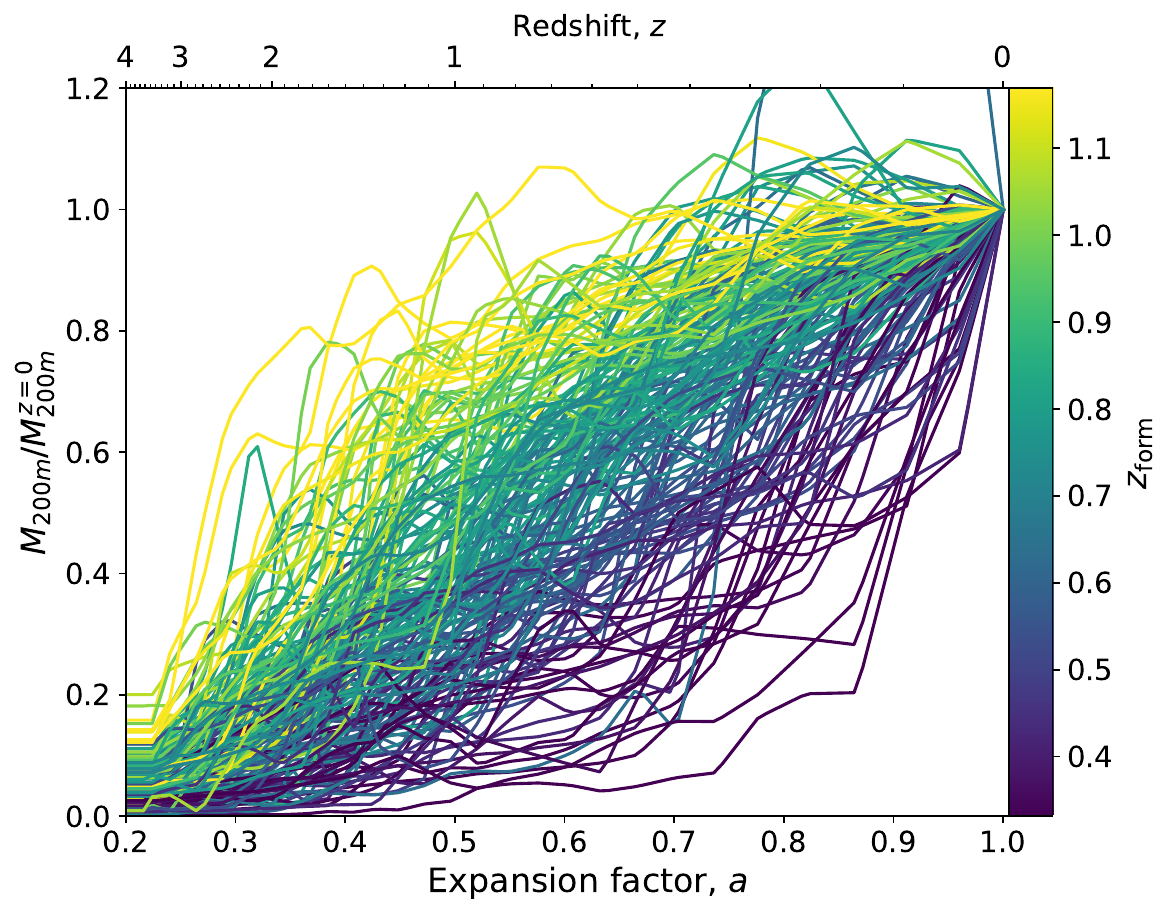}~
        \includegraphics[width=6cm]{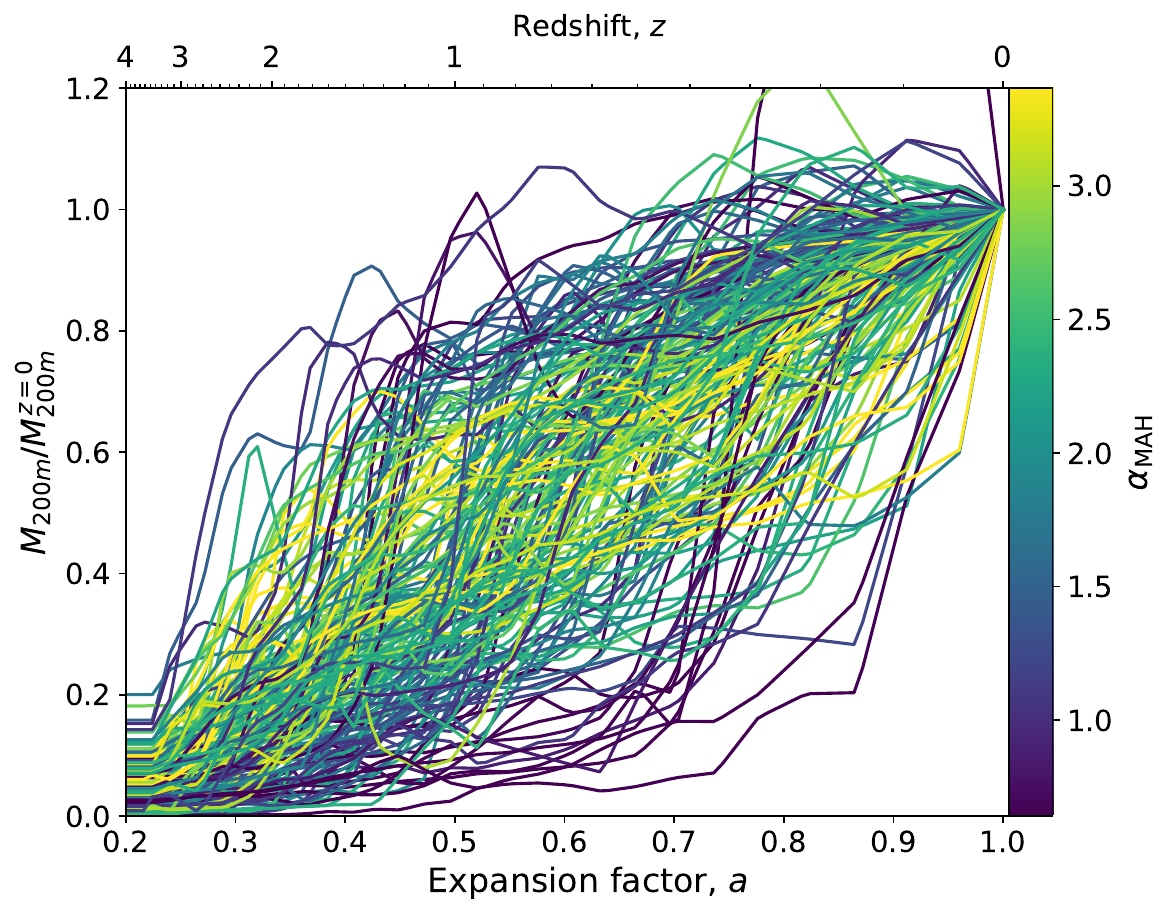}\\
        {\hspace{-0.04\textwidth}
        \includegraphics[width=6cm]{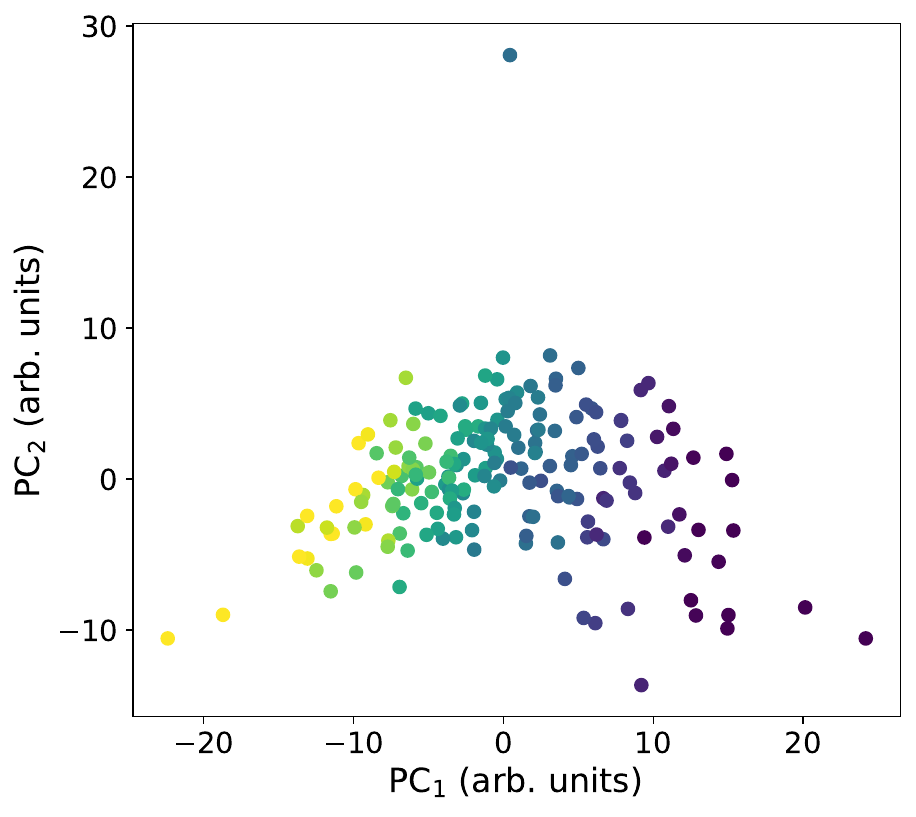}~
        \includegraphics[width=6cm]{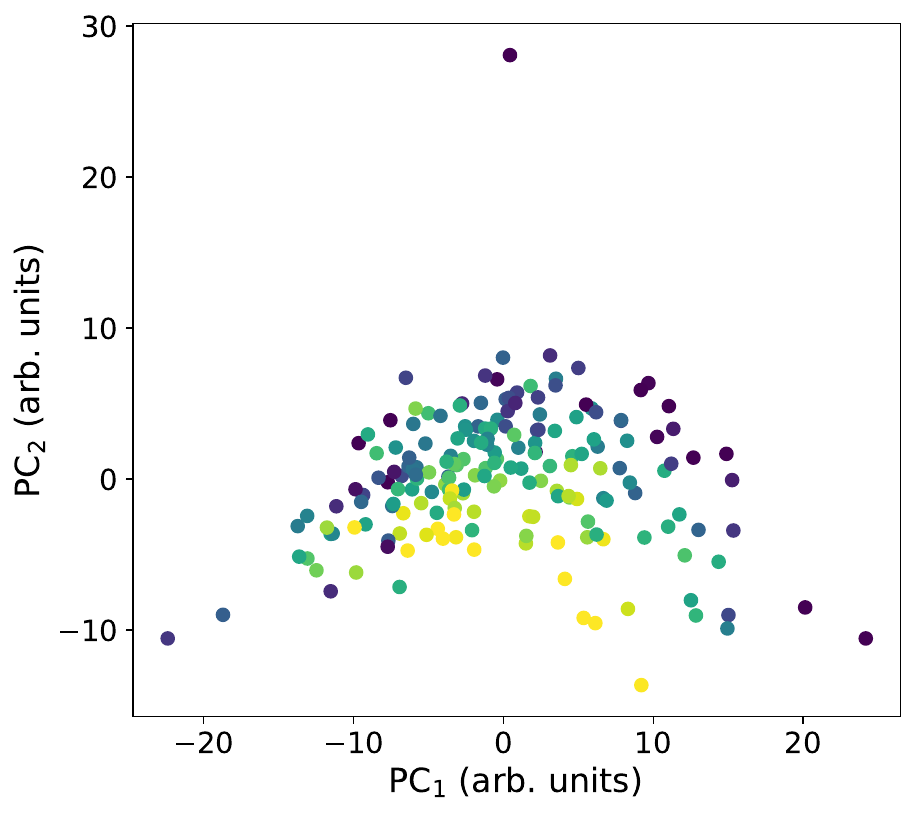}
        }
    \end{minipage}\hfill
    \begin{minipage}[t]{0.28\textwidth}
        \caption{Parametrisation of the MAHs. The two top panels present the assembly histories of the whole sample, colour-coded according to the formation redshift $z_\mathrm{form}$ (left) and the episodicity parameter $\alpha_\mathrm{MAH}$ (right-hand side panel). The two bottom panels describe the distribution of these parameters in the $\{\mathrm{PC}_1, \, \mathrm{PC}_2 \}$ space, accounting for the maximum variance with two linearly uncorrelated components. The colour coding of the dots is the same as represented in their respective top panels.}
        \label{fig:MAH_zform_alpha_PCA}
    \end{minipage}
\end{figure*}

In the two upper panels of Fig.~\ref{fig:MAH_zform_alpha_PCA}, we plot all MAHs, normalised as $M_{200m}(a) / M_{200m}(a_0=1)$, colour-coded by $z_\mathrm{form}$ (left-hand side panel) and by $\alpha_\mathrm{MAH}$ (right-hand side panel). 
The interpretation of the former is direct, earlier-assembling clusters having larger --yellow-- values of $z_\mathrm{form}$ (in contrast, later-assembling clusters having smaller --purple-- values).
Clusters with large $\alpha_\mathrm{MAH}$ (yellow lines) have straighter MAHs, indicative of a smoother --at a more constant pace-- assembly of their mass. Conversely, the smaller the value of $\alpha_\mathrm{MAH}$ (turquoise to purple lines), the larger the variation in the steepness of these curves, presenting a very steep build-up of mass around $z_\mathrm{form}$ (coming from one or several successive major mergers) and a quiescent phase for the rest of their assembly history.

While this parametrisation is well physically-motivated, it is still imperative to check to what extent does it describe the diversity of assembly histories. To this end, we have performed a PCA on our MAHs, qualitatively reproducing the results by \citet{Wong_2012} for the two first components, jointly accounting for $77\%$ of their total variance (cf. $83\%$ in the aforementioned reference; for a thorough description, see Appendix~\ref{s:app.PCA}).

In the two lower panels of Fig.~\ref{fig:MAH_zform_alpha_PCA}, we plot the distribution of our assembly histories in the $\{\mathrm{PC}_1, \mathrm{PC}_2\}$ space, colour-coded according to $z_\mathrm{form}$ (left) and $\alpha_\mathrm{MAH}$ (right) following the colour bars in the respective panels above. There is a very clean correlation (albeit non-linear) between $z_\mathrm{form}$ and $\mathrm{PC}_1$, resulting in a Spearman rank correlation $\rho_\mathrm{sp}(z_\mathrm{form}, \, \mathrm{PC}_1) = - 0.975$, justifying our choice in Sect.~\ref{s:results.MAHs_vs_dynstate} to characterise the dependence of $z_\mathrm{form}$ on our indicators, as the main summary statistic of the full MAH.

For $\alpha_\mathrm{MAH}$, from the bottom right panel of Fig.~\ref{fig:MAH_zform_alpha_PCA} it is obvious that, at fixed $\mathrm{PC}_1$, $\mathrm{PC}_2$ is significantly anticorrelated with $\alpha_\mathrm{MAH}$. To quantify the strength of such correlation, since data points in the $\{ \mathrm{PC}_1, \, \mathrm{PC}_2 \}$ are distributed in an arc-like shape (hence, non-monotonic), we use the non-parametric residual correlation introduced in Sect.~\ref{s:methods.statistics.partial}. In this way, we obtain that the residuals with respect to the underlying $\mathrm{PC}_2 = f (\mathrm{PC}_1)$ relation have a Spearman rank correlation of $-0.68$ with $\alpha_\mathrm{MAH}$ (up to $-0.72$ if discarding the most extreme outliers, $|\mathrm{PC}_1|>10$). This implies that our set of two physically-motivated variables describe, to a considerably large extent, the main variability of the assembly histories in our sample as revealed by the PCA, and therefore are in principle a good choice to study the dependence of the MAHs on assembly state indicators at $z=0$ below.

\subsubsection{Constraints on the MAHs from the assembly state indicators}
\label{s:results.MAHs_biparametric.constraints}

Having defined a two-dimensional space where the complexity of MAHs can be summarised, we can now investigate what direction in this space is constrained by each of the indicators in Sect.~\ref{s:methods.indicators}. We do so under the assumption of monotonous, but not necessarily linear, dependences of our parameters with the MAHs within this space, which is well justified based on the findings of the previous Sect.~\ref{s:results.MAHs_vs_dynstate}.

\begin{figure*}
    \begin{minipage}[t]{0.7\textwidth}
    \centering
    \includegraphics[width=10cm]{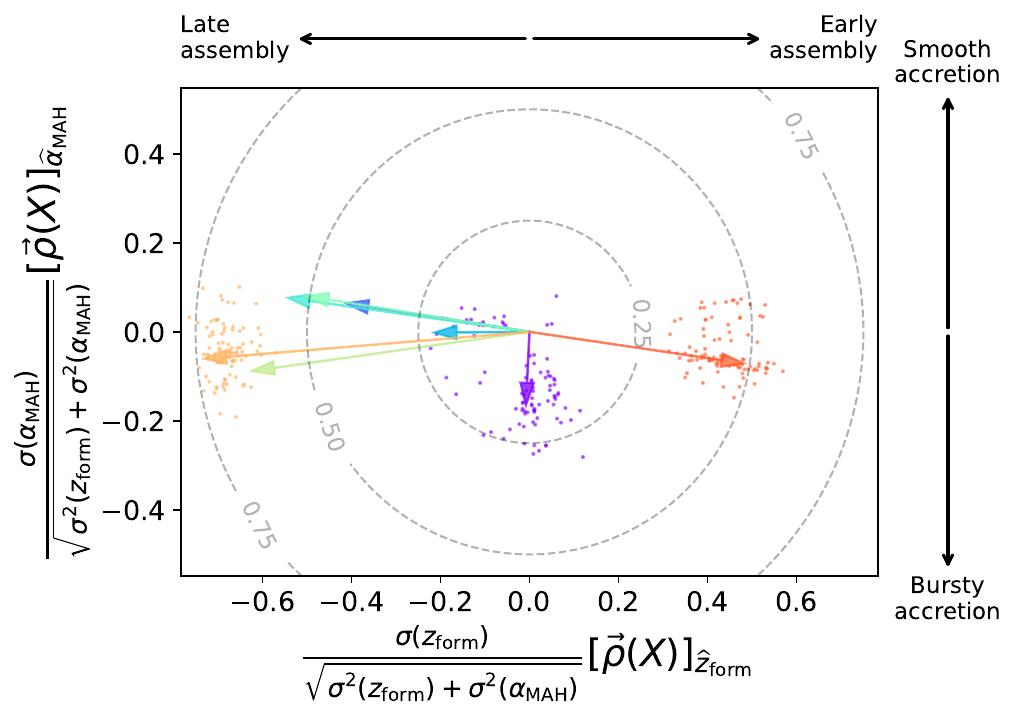} \\
    \includegraphics[width=10cm]{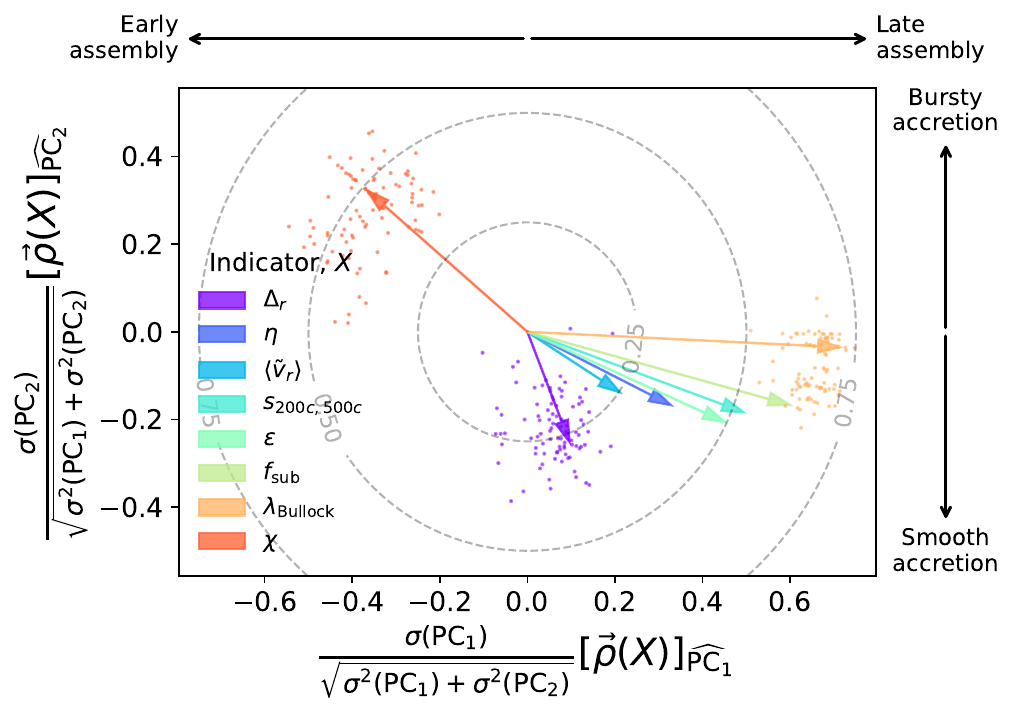}
    \end{minipage}\hfill
    \begin{minipage}[t]{0.28\textwidth}
    \caption{Constraints of the assembly state indicators on the two biparametric characterisations of the MAHs: the ${z_\mathrm{form} - \alpha_\mathrm{MAH}}$ plane (top panel) and the ${\mathrm{PC}_1 - \mathrm{PC}_2}$ plane (bottom panel). In each panel, the different arrows correspond to each of the assembly indicators according to the colour legend in the bottom panel. Their directions indicate the relative weight of each of the two parameters in the linear combination which best correlates with the given indicator, while their magnitude is the Spearman rank correlation coefficient between the indicator and the corresponding projection of the MAH representation (the larger, the more information is yielded by the indicator). \\
    \textit{\underline{Colour aid:} in  the top panel, clockwise from above, the arrows correspond to $\chi$, $\Delta_r$, $f_\mathrm{sub}$, $\lambda_\mathrm{Bullock}$, $\langle \tilde{v}_r \rangle$, $s_{200c,500c}$, $\eta$, and $\epsilon$. Likewise, in the bottom panel, clockwise from above, the arrows are $\lambda_\mathrm{Bullock}$, $f_\mathrm{sub}$, $s_{200c,500c}$, $\epsilon$, $\eta$, $\langle \tilde{v}_r \rangle$, $\Delta_r$, and $\chi$.}
    }
    \label{fig:variation_directions}
    \end{minipage}
\end{figure*}

Following the procedure of Sect.~\ref{s:methods.statistics.direction}, for each of the parameters, $X$, at $z=0$ we can obtain a vector $\vec{\rho}(X)$ representing the direction in which $X$ varies, and the magnitude of such dependence (the larger it is, the more information about the MAH that the parameter can yield). This is represented, both in the ${z_\mathrm{form} - \alpha_\mathrm{MAH}}$ and in the ${\mathrm{PC}_1 - \mathrm{PC}_2}$ planes, in the upper and lower panels of Fig.~\ref{fig:variation_directions}, respectively.

In the former case, i.e. in the ${z_\mathrm{form} - \alpha_\mathrm{MAH}}$ plane, it appears that little information can be obtained on the episodicity of accretion from this set of parameters (vectors are mostly horizontal). Only $\Delta_r$ provides very marginal ($p$-value $= 0.11$) correlation with $ \alpha_\mathrm{MAH}$, while the only relevant information being drawn from the remaining of parameters lies in the early-late assembly axis, i.e., corresponds to $z_\mathrm{form}$. Naturally, our combined indicator $\chi$ points in the opposite direction to the inidivdual ones, since it measures relaxedness, as opposed to the rest which are intuitively measures of disturbance. Strikingly, amongst these parameters, it is the spin parameter the one most correlated with the MAH (mainly, $z_\mathrm{form}$), achieving a Spearman rank correlation of $0.68$.

Perhaps more interestingly, if we look at these relations in the $\mathrm{PC}_1 - \mathrm{PC}_2$ plane, there is a broader separation of these vectors, indicating that one can obtain broader information on these parameters from the assembly state indicators. In particular, general trends are maintained (e.g., the spin parameter yielding the most amount of information, $\chi$ indicating relaxedness and hence being opposed to the rest of indicators, etc.). However, there is a greater scope of the range of information about the MAH yielded by each parameter. For instance, while halo spin informs exclusively about formation time (larger spin implying later assembly), larger centre offset is indicative of smoother accretion (less episodic) regardless of formation time. Other indicators (substructure fraction, sparsity, ellipticity, virial ratio and mean radial velocity) lie in between these extreme cases, encompassing increasingly more information about $\mathrm{PC}_2$ in the order specified above.

Despite the low significance of the results in the ${z_\mathrm{form} - \alpha_\mathrm{MAH}}$ plane, it might be striking how $\Delta_r$ appears as an indicator of \textit{bursty} accretion when assessed in the ${z_\mathrm{form} - \alpha_\mathrm{MAH}}$ space, while it indicates smoother accretion in the principal components space. These results are not necessarily at odds with each other, but rather reflect that $\mathrm{PC}_2$ and $\alpha_\mathrm{MAH}$ (contrary to $\mathrm{PC}_1$ and $z_\mathrm{form}$) are correlated but not fully codependent. Thus, even though we have dubbed both the regimes of high $\mathrm{PC}_2$ and low $\mathrm{\alpha_\mathrm{MAH}}$ as ``\textit{bursty}'' accretion, they represent episodicity in different senses. While $\alpha_\mathrm{MAH}$ represents a mean accretion rate in a long and variable time interval (between the instant when $25\%$ and $75\%$ of the present-day mass was built up), $\mathrm{PC}_2$ is focused at a particular redshift (around $a \simeq 0.65$; see Fig. \ref{fig:PCA_real_appendix}). Additionally, despite the fact that $\alpha_\mathrm{MAH}$ appears to be more physically-motivated, it is reasonable that we see more significant relations with $\mathrm{PC}_2$, since --by construction-- the latter variable captures more variance of the MAHs than the former.

\begin{figure*}
    \centering 
    \includegraphics[width=.8\linewidth]{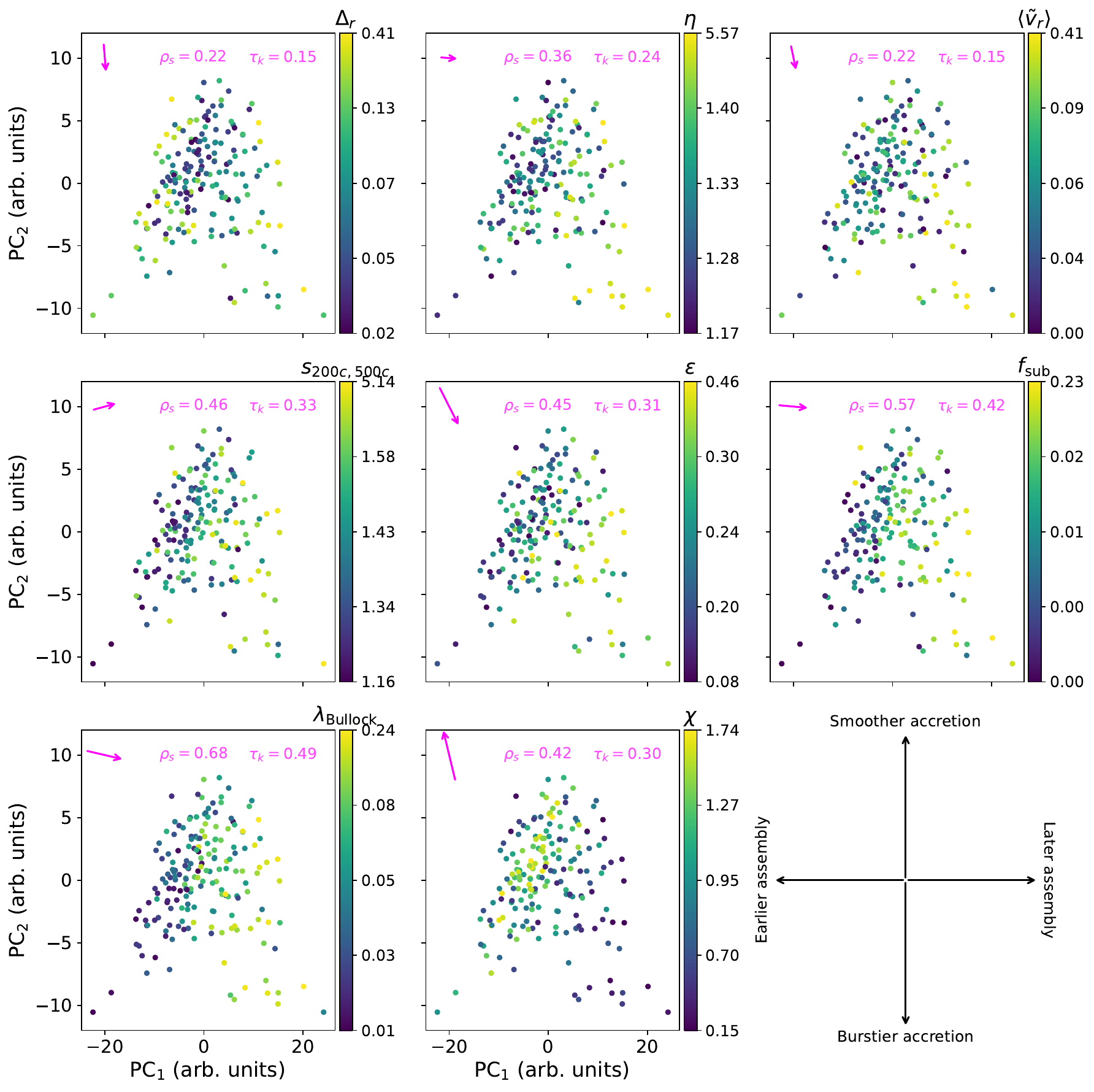}
    \caption{Assembly histories for individual clusters (summarised in the $\mathrm{PC}_1 - \mathrm{PC}_2$ plane) colour-coded according to each of the assembly state indicators at $z=0$. The colour bars are ordinal (i.e., linear in the ranked variable: e.g., the middle point of the colorbar corresponds to the median value). At the top of each panel, we indicate the direction of maximum correlation with the indicator, the Spearman rank correlation coefficient, and the Kendall rank correlation coefficient.}
    \label{fig:scatter_PC1PC2}
\end{figure*}

This is why, in Fig. \ref{fig:scatter_PC1PC2}, we have focused on the $\mathrm{PC}_1 - \mathrm{PC}_2$ space to represent the diversity of assembly histories in our sample of haloes, colour-coded according to each of the indicators. Generally, trends are difficult to identify visually due to the considerable scatter. This is why, in the top part of each panel, we have annotated the direction of maximum correlation with the variable represented in colour, the magnitude of such correlation, and also the Kendall rank correlation coefficient (see Sect. \ref{s:methods.statistics.interpretation} for the definitions). Thus, the arrows and Spearman coefficients present identical information to what has been shown in the lower panel of Fig. \ref{fig:variation_directions}, but the scatter plot and $\tau_K$ values give a clearer view on how much information is brought by each single parameter.

In particular, a probabilistic interpretation of $\tau_K (x, y)$ comes by noting that the probability that, if $x_1 \geq x_2$, then $y_1 \geq y_2$, is precisely $(1 + \tau_K)/2$. Therefore, the most promising indicator for constraining the past formation history of galaxy clusters and groups is, consistent with previous results, $\lambda_\mathrm{Bullock}$. In this case, finding a cluster with higher spin parameter than another one gives $75\%$ probability that it will be later forming (more precisely, it will have a larger component along the direction specified by the arrow in the bottom, left panel of Fig. \ref{fig:scatter_PC1PC2}, which is essentially the $z_\mathrm{form}$ axis).

Other parameters ($f_\mathrm{sub}$, $s_{200c,500c}$, etc.) still provide significant constraints on the position of the cluster in this assembly history representation space, while in other cases ($\Delta_r$, $\langle \tilde v_r \rangle$) the information on a cluster-to-cluster basis is marginal ($\tau_K \lesssim 0.2$, implying probabilities of concordant pairs below $60\%$, which is only marginally better than random guessing).

Therefore, while on an ensemble sense these parameters predict very noticeable segregation of the MAH curves and derived parameters (see Sect. \ref{s:results.MAHs_vs_dynstate}), what could be useful for e.g. sample selection, on a one-to-one basis the population scatter is sufficiently large to significantly worsen the prospects of determining the assembly history of a halo by looking at a single parameter. Nevertheless, as we have shown, different parameters provide constraints in different directions in this space that summarises the MAHs. Hence, this leaves the door open to the combination of different indicators to better constrain the location of objects in this space.

\subsection{Assembly state inferred from the ICM distribution}
\label{s:results.gas_vs_DM}

All the quantities involved in the study are derived from the dark matter distribution. As a first step towards the connection of these quantities to actually observable data, in this section we explore how much of this information is contained in the ICM morphology and dynamics. While the ICM properties defined here are not directly observable, a tight correlation between three-dimensional ICM and DM properties is a necessary (but not sufficient) condition for SZ and X-ray morphology to be capable of constraining assembly state. In Sect. \ref{s:results.gas_vs_DM.definitions} we describe how we compute these parameters for the gas distribution, while the results of the comparison are discussed in Sect. \ref{s:results.gas_vs_DM.results}.

\subsubsection{Definition of the indicators for the ICM}
\label{s:results.gas_vs_DM.definitions}
In order to define an equivalent set of indicators computed from the ICM (again using the whole, three-dimensional description), we need to modify the definition or the computation process of several of the parameters. In particular,

\noindent \textbf{Centre offset.} The computation of the centre of mass is analogous to the DM case. For the gas density peak, however, instead of following the procedure described in \citet{Valles-Perez_2022} for DM particles, we take advantage of the AMR grid. Thus, we iterate from coarser to finer levels in the AMR hierarchy, each time finding the density peak in a $3 \times 3 \times 3$-cells neighbourhood of the location determined at the previous AMR level. This iteration is especially important in our case, since the usage of simulations with cooling but no feedback can produce a significant number of unphysical clumps away from the cluster centre.

\noindent \textbf{Virial ratio.} Due to the collisional nature of gas, most of its kinetic energy ($T$) is converted into internal ($U_\mathrm{th}$). Thus, we consider the quantity $T' = (T+U_\mathrm{th}) / f_\mathrm{B}$, where $f_\mathrm{B} = 0.155$ is the cosmic baryon fraction, under the assumption that the total kinetic energy can be extrapolated to the whole mass distribution. Similarly, from the gravitational binding energy of the gas to itself ($U_\mathrm{gas}$), we estimate $U' = U_\mathrm{gas} / f_\mathrm{gas}^2$.

\noindent \textbf{Sparsity.} In this case, sparsity is computed as by finding $M_{\Delta_{c,\text{gas}}}$ as the mass within a sphere containing a mean gas overdensity $\Delta_{\text{gas}} = f_\mathrm{gas} \Delta_c $ with respect to the critical $\Lambda$CDM density.

\noindent \textbf{Substructure fraction.} The notion of substructure is fundamentally different for the ICM and for the DM halo. Many DM substructures do not have a corresponding gaseous halo, while the ICM contains many clumps not associated with a DM substructure. This is why we have opted to quantify gaseous substructure in its most natural way, using a method based on \citet{Zhuravleva_2013}. Essentially, we take directional profiles of the ICM density with $\Delta \log r = 0.01 \, \mathrm{dex}$ and, at each $r$, we flag as gaseous substructure (i.e., clumps) all volume elements exceeding a density $\log \rho = \log \tilde \rho + f_\mathrm{cut} \sigma_{\log \rho}$, where $\tilde \rho$ is the median density, $\sigma_{\log \rho}$ is the standard deviation of $\log \rho$ and $f_\mathrm{cut}=3.5$ is a free parameter, for which we take the same prescription as \citet{Zhuravleva_2013}.

The rest of indicators not mentioned here are computed in the same way as for DM. As for our combined indicator $\chi$, since it corresponds to an empirical calibration of thresholds and weights, it lacks sense applying it directly to other indicators defined from the gas distribution. We note the reader that the dynamics of the ICM is strongly affected by galaxy formation physics and other physical effects (e.g., viscosity) not considered in this simulation. Thus, one can expect shifts in the values of these indicators (e.g., feedback may suppress clumping, thus reducing substructure fraction from gas). The simpler physical setup in this simulation lets us isolate the effects of structure formation driven phenomena.

\subsubsection{Assembly state indicators from the ICM}
\label{s:results.gas_vs_DM.results}

\begin{figure*}
    \centering
    \includegraphics[width=0.3\textwidth]{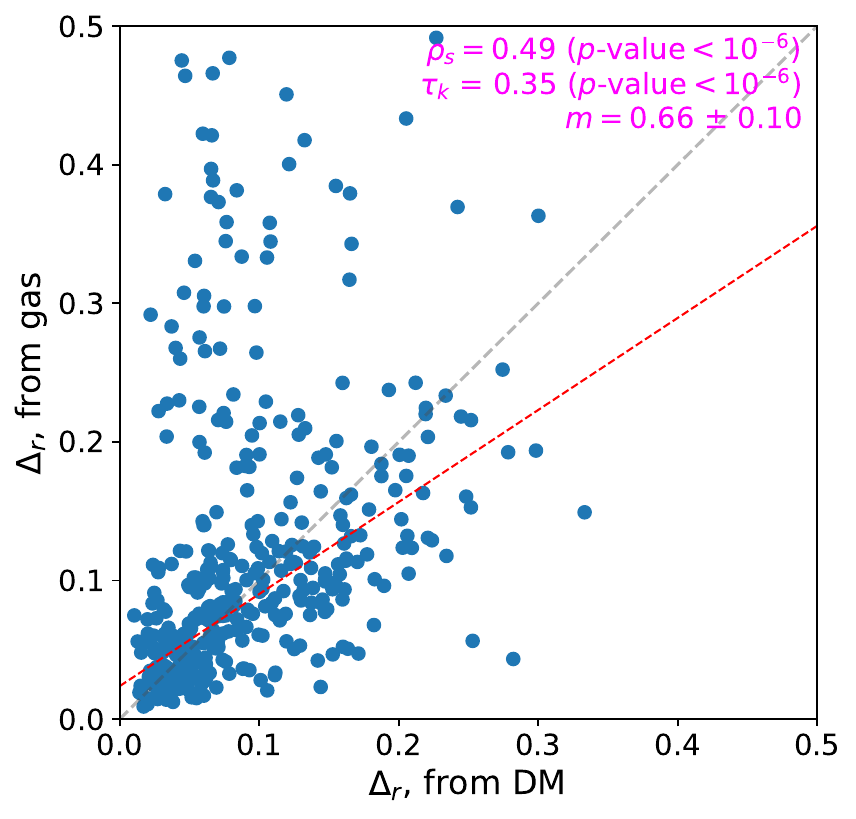}~
    \includegraphics[width=0.3\textwidth]{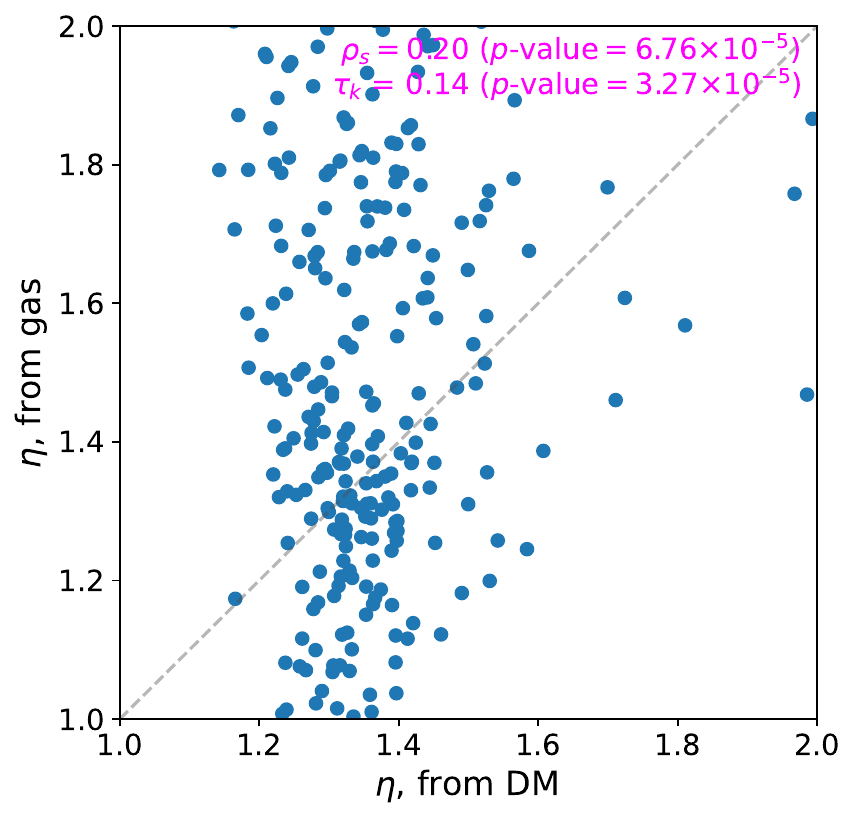}~
    \includegraphics[width=0.3\textwidth]{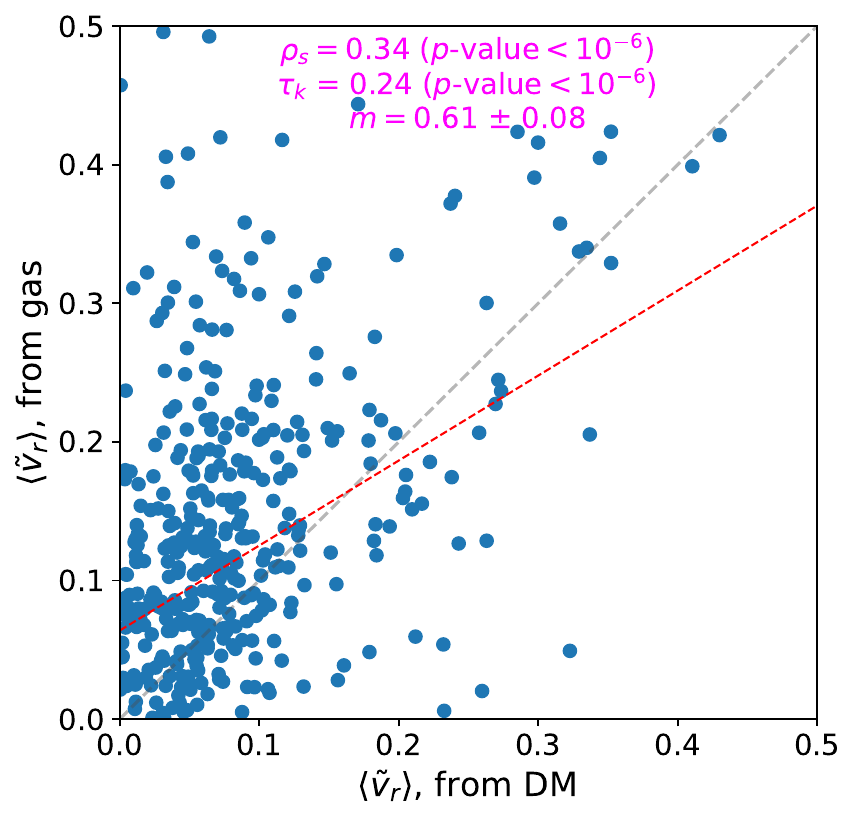}
    \includegraphics[width=0.3\textwidth]{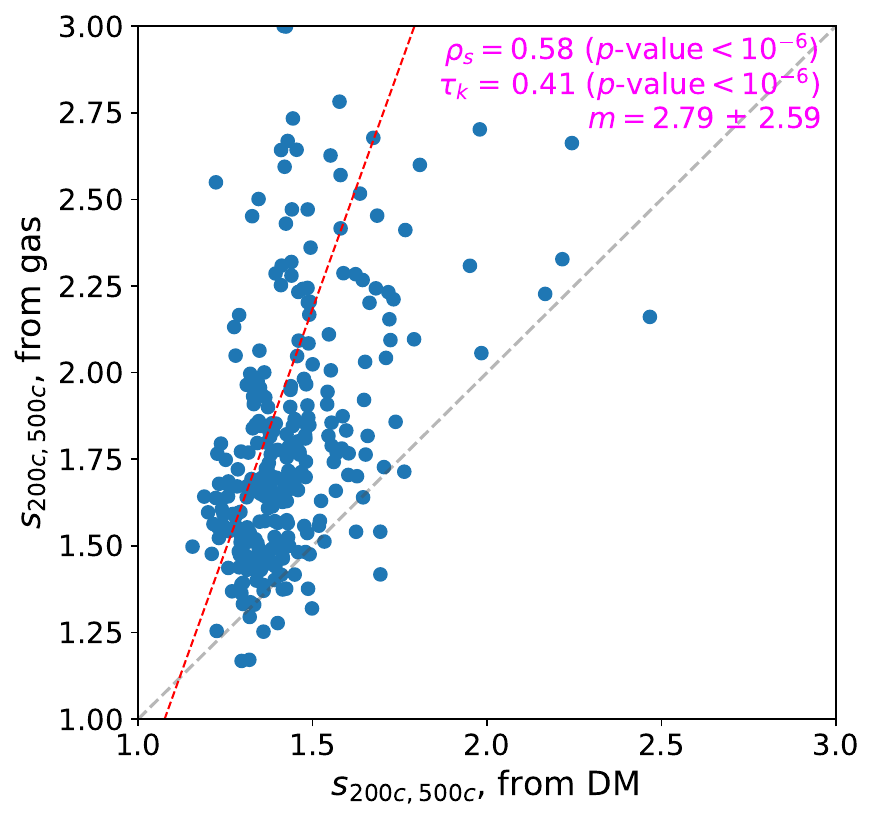}~
    \includegraphics[width=0.3\textwidth]{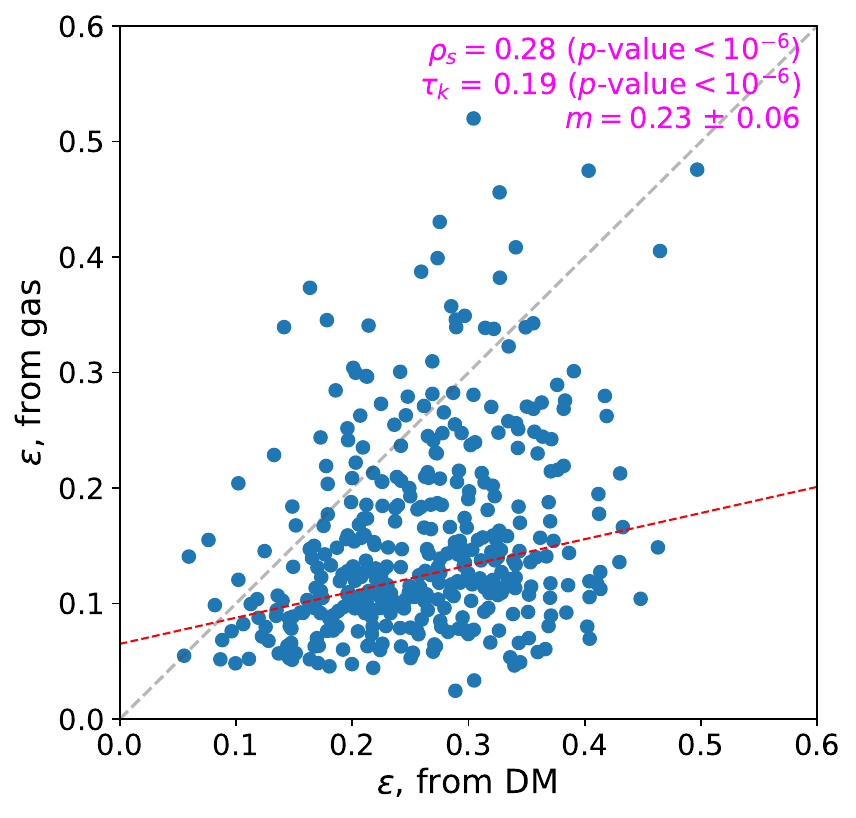}~
    \includegraphics[width=0.3\textwidth]{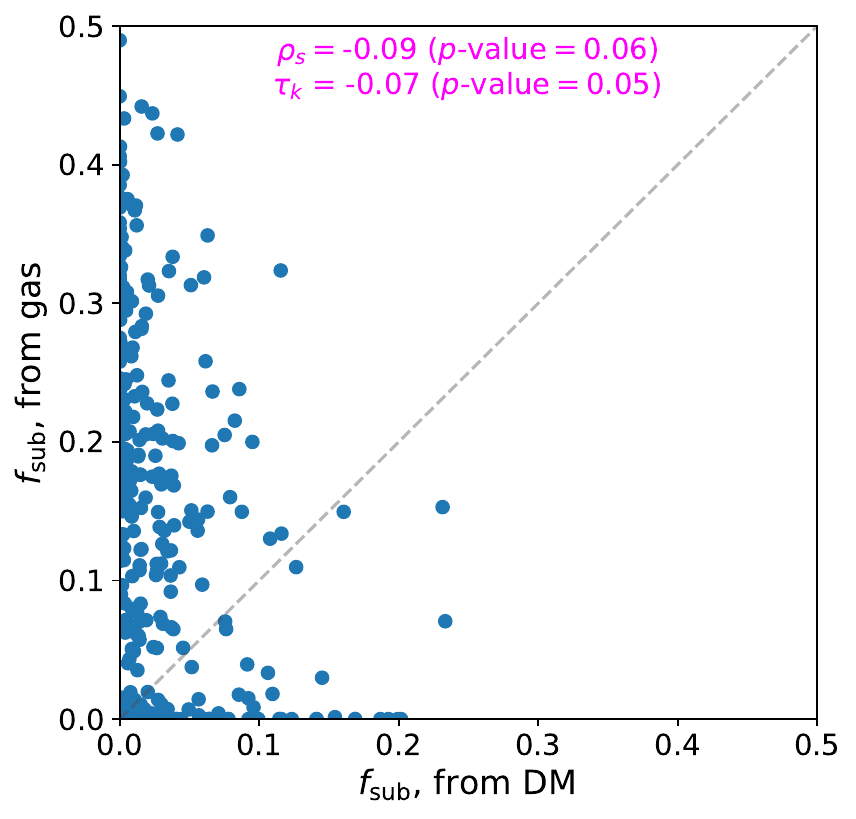}
    \includegraphics[width=0.3\textwidth]{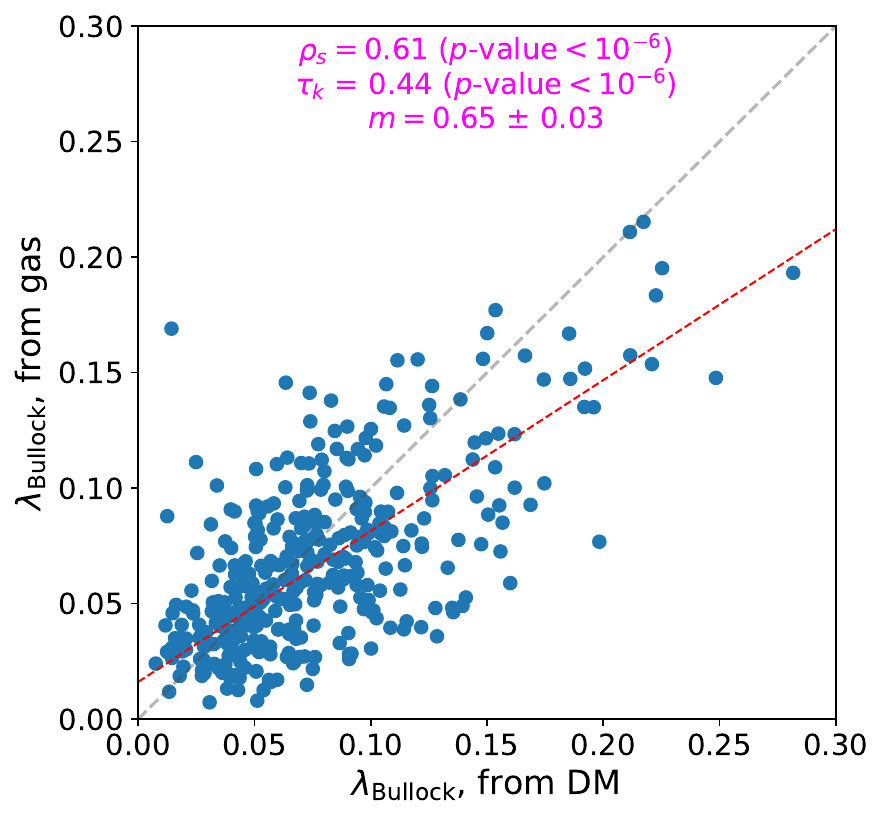}
    \caption{Correspondence between the assembly state indicators measured from the DM distribution (horizontal axes) and from the ICM distribution (vertical axes), at $z=0$. Each panel corresponds to a different parameter. Within each panel, each dot corresponds to a cluster or group; the dashed, gray line is the identity relation ($X^\mathrm{gas} = X^\mathrm{DM}$); and the dashed, red line is a least-squares linear fit, whose slope ($m$) is given at the top of each panel. Also the Spearman and Kendall rank correlation magnitudes, and their $p$-values, are given there.}
    \label{fig:DM_vs_gas}
\end{figure*}

The results of the comparison between the set of indicators computed from the DM distribution, and from the ICM distribution, are summarised in Fig. \ref{fig:DM_vs_gas}. In general, there are positive correlations for most of the parameters, but the scatter is sufficiently high to prevent a clear measurement of DM halo properties from ICM properties.

In many cases ($\Delta_r$, $\langle \tilde v_r \rangle$, $\varepsilon$, and $\lambda_\mathrm{Bullock}$), DM and gas indicators show a moderate ($\rho_s \sim 0.3 - 0.6$) correlation, with very significant scatter, and a slope $m < 1$ (represented with the dashed, red line), clearly inconsistent with a direct one-to-one relation (represented by the dashed, gray line). This is most likely associated to the collisional nature of the ICM, which tends to produce rounder shapes \citep{Lau_2011, Lau_2012}, with consequently lower $\Delta_r$, and lower infall velocities.

For these indicators (with the exception of $\lambda_\mathrm{Bullock}$), there are strong outliers above the fitted relation, indicating the presence of objects with very disturbed gas morphology, compared to their DM distribution. It is plausible that the presence of unopposed cooling in this simulation is generating some of these very disturbed morphologies, although fully ellucidating their origin falls beyond the scope of the present work. 

Amongst all these parameters, it is $\lambda_\mathrm{Bullock}$ the one yielding the relation with the least scatter. This is interesting, since this quantity also presents the largest correlations with assembly history (see Sect. \ref{s:results.MAHs_biparametric}). Additionally, this quantity could, potentially --and only in projection--, be determined by kinetic Sunyaev-Zeldovich measurements (\citealp{Baldi_2018}; albeit with large scatter, cf. \citealp{Monllor_2024}), rendering it a useful indicator to further study in both theoretical and observational data.

Other cases present even more striking differences between gas and DM. In particular, $\eta$ values are essentially uncorrelated amongst the two components, signalling that the naive corrections by $f_\mathrm{gas}$ suggested in Sect. \ref{s:results.gas_vs_DM.definitions} are not effective. To some degree, this could be contributed by the departure of $f_\mathrm{gas}$ from its cosmic average (used for the corrections on $T'$ and $U'$). However, this simulation lacks feedback and $f_\mathrm{gas}$ values present a rather narrow distribution, only slightly below the cosmic value. Instead, the most likely explanation for the lack of correlation resides in the different density profiles of both components, which make the comparison between gravitational energies inferred from the gas or from the total mass not straightforward. Perhaps less surprising is the lack of correlation between substructure fractions. Generally, we find cases of, both, rich DM substructure with no gaseous substructure and, more abundantly, the opposite. This is reasonable since these substructures have fundamentally different origins: the DM one comes from accretion of smaller haloes, while the gaseous one (clumpiness) occurs in-situ due to cooling, even though it has also been linked to different measures of dynamical unrelaxedness \citep{Roncarelli_2013, Vazza_2013, Battaglia_2015, Planelles_2017}.

Lastly, the case of sparsity is interesting, since it presents a moderate correlation with slope $m > 1$. That is to say, while clusters appear \textit{more relaxed} in their gaseous component when using indicators such as $\Delta_r$, $\varepsilon$, etc., the opposite trend happens for $s_{200c,500c}$. As a matter of fact, gas sparsity appears to be approximately lower-bounded by DM sparsity, as a result of the gas distribution usually being less concentrated. 

To sum up, while in some cases DM and gas indicators present moderate correlations, the determination of these indicators from ICM data appears to be complex and affected by high scatter (let aside projection effects). On top of that, these correlations are not scattered around the identity, but rather the slope can be larger or smaller than unity, so care must be taken when comparing relaxation criteria amongst different components.

\section{Discussion and conclusions}
\label{s:conclusions}

In this work, we have used a moderate-sized hydrodynamical + $N$-Body cosmological simulation to study the connection between DM halo and ICM properties and the full assembly history of a sample of galaxy clusters and groups ($10^{13} M_\odot \leq M_\mathrm{vir} \lesssim 8 \times 10^{14} M_\odot$) through the redshift interval $4 \geq z \geq 0$. The extreme diversity in the formation channels of these structures demands a careful study of how these complex processes shape their properties, on what timescales do they imprint them, and what is the extent of such an effect.

\begin{table*}[]
    \centering
    
    \definecolor{earliest}{rgb}{0.1, 0.2, 0.6} 
    \definecolor{early}{rgb}{0.2, 0.4, 0.7}    
    \definecolor{mid}{rgb}{0.4, 0.6, 0.8}      
    \definecolor{late}{rgb}{0.45, 0.675, 0.9}     

    \definecolor{lowest}{rgb}{0.8, 0.2, 0.2}      
    \definecolor{low}{rgb}{0.8, 0.5, 0.2} 
    \definecolor{medium}{rgb}{0.7, 0.7, 0.2}   
    \definecolor{high}{rgb}{0.2, 0.8, 0.2}     

    \definecolor{onlyepisodicity}{rgb}{0.0, 0.8, 0.8} 
    \definecolor{mostlyepisodicity}{rgb}{0.3, 0.6, 0.7} 
    \definecolor{both}{rgb}{0.5, 0.5, 0.5} 
    \definecolor{mostlytiming}{rgb}{0.7, 0.3, 0.7} 
    \definecolor{onlytiming}{rgb}{0.8, 0.0, 0.8} 

    \caption{Results summary for the timing analysis (central block) and the biparametric analysis (right-hand side block) on the MAHs, indicator by indicator in each row. With respect to timing analysis, the first column specifies whether the indicator responds to instantaneous accretion (\textcolor{late}{\textit{late-time}}), accretion $\sim \tau_\mathrm{dyn}/4$ ago (\textcolor{mid}{\textit{mid-time}}), $\sim \tau_\mathrm{dyn}/2$ ago (\textcolor{early}{\textit{early-time}}) or higher (\textcolor{earliest}{\textit{earliest-time}}). The subsequent two columns categorically assess the strength of the maximum correlation between the indicator and the accretion rates $\Gamma_{200m}(z_c)$, as \textcolor{lowest}{\textit{lowest}} (correlation $\sim 0.2$), \textcolor{low}{\textit{low}} ($\sim 0.3$), \textcolor{medium}{\textit{medium}} ($\sim 0.4$), and \textcolor{high}{\textit{high}} ($\sim 0.5$), when the indicators are determined at $z=0$ and at $z=0.5$, respectively. Regarding the biparametric analysis (accretion timing and episodicity), we specify whether each parameter informs about \textcolor{mostlytiming}{mostly}/\textcolor{onlytiming}{only timing} or \textcolor{mostlyepisodicity}{mostly}/\textcolor{onlyepisodicity}{only episodicity}, or a \textcolor{both}{mixture of both}. The second column categorises the strength of the correlation with the whole MAH component in \textcolor{low}{\textit{low}} ($\sim 0.25$), \textcolor{medium}{\textit{medium}} ($\sim 0.5$) and \textcolor{high}{\textit{high}} (up to $\sim 0.75$).}
    \begin{tabular}{c||c|cc||c|c}
        & \multicolumn{3}{c||}{Timing analysis (\S \ref{s:results.MAHs_vs_dynstate})} & \multicolumn{2}{|c}{Episodicity vs. timing analysis (\S \ref{s:results.MAHs_biparametric})} \\ \hline
        & \multirow{2}{*}{Timing} & \multicolumn{2}{|c||}{(strength)} & Feature being &  \\
        & & at $z=0$ & at $z=0.5$ & constrained & (strength) \\
        \hline \hline
        
         $\Delta_r$ & \textcolor{mid}{mid-time} & \textcolor{lowest}{lowest} & \textcolor{lowest}{lowest} & \textcolor{onlyepisodicity}{only episodicity} & \textcolor{low}{low} \\
         $\eta$ & \textcolor{mid}{mid-time} & \textcolor{medium}{medium} & \textcolor{low}{low} & \textcolor{mostlytiming}{mostly timing} & \textcolor{low}{low}\\
         $\langle \tilde v_r \rangle$ & \textcolor{late}{late-time} & \textcolor{medium}{medium} & \textcolor{high}{high} & \textcolor{both}{both} & \textcolor{low}{low}\\ 
         $s_{200c,500c}$ & \textcolor{earliest}{earliest time} & \textcolor{medium}{medium} & \textcolor{lowest}{lowest} & \textcolor{mostlytiming}{mostly timing} & \textcolor{medium}{medium}\\ 
         $\epsilon$ & \textcolor{early}{early-time} & \textcolor{medium}{medium} & \textcolor{medium}{medium} & \textcolor{mostlytiming}{mostly timing} & \textcolor{medium}{medium}\\ 
         $f_\mathrm{sub}$ & \textcolor{early}{early-time} & \textcolor{high}{high} & \textcolor{high}{high} & \textcolor{mostlytiming}{mostly timing} & \textcolor{high}{high}\\ 
         $\lambda_\mathrm{Bullock}$ & \textcolor{early}{early time} & \textcolor{medium}{medium} & \textcolor{medium}{medium} & \textcolor{onlytiming}{only timing} & \textcolor{high}{high}\\ 
         $\chi$ & \textcolor{mid}{mid-time} & \textcolor{high}{high} & \textcolor{high}{high} & \textcolor{both}{both} & \textcolor{medium}{medium}
    \end{tabular}
    \label{tab:summary}
\end{table*}

The main findings of this paper can be summarised and put into context within the bibliography as follows:

\begin{enumerate}
\itemsep .5em

\item The strategy for building the merger tree and, especially, selecting its main branch, is critical for determining and comparing the MAHs. In particular, at $z \simeq 4$, ours (where the main branch of the merger tree is determined on the basis of gravitational binding) lie a factor of $\sim 3$ [$\sim 2$] below the ones of \citet{Correa_2015} [\citealp{vandenBosch_2014}], both of which are based on selecting the most massive progenitor.\\[-.8em]

While there is not a `true' solution for defining the main branch of the merger tree, it is important to take caution when comparing amongst works in the literature, since different choices lead to noticeable differences. In particular, in Appendix \ref{s:app.mtree} we show how, if we switch our strategy for defining the main branch, our MAHs are brought to substantially better agreement with the previously mentioned works.

\item As already discussed in different forms by \citet{Jeeson-Danie_2011}, \citet{Skibba_2011} and \citet{Haggar_2024}, indicators for the dynamical state of a cosmic structure (either galaxy clusters or their DM haloes) are rather loosely correlated. In Fig. \ref{fig:corner} we have shown how only a few pairs of parameters reach Spearman correlation ranks above $\rho_s \sim 0.4$. This already suggests that \textit{(i)} the information brought on by a single indicator can be fairly limited in a cluster-by-cluster basis; and \textit{(ii)} combinations of parameters might be useful to improve the predictive power on the MAHs, as done for instance by \citetalias{Valles-Perez_2023}.

\item Selecting clusters on the assembly state indicators defined in Sect. \ref{s:methods.indicators} clearly splits their ensemble (stacked) MAHs in such a way that, in all cases, more disturbed haloes according to any given parameter at $z=0$ are more late-forming (Fig. \ref{fig:MAH_parameters}).
\begin{enumerate}
    \item Upon closer inspection (Fig. \ref{fig:MAH_spearman}), these quantities are especially informative about different moments of the MAHs, enabling us to categorise them into more late-time (e.g., $\langle \tilde v_r \rangle$, $\eta$, etc.) and more early-time (e.g., $\epsilon$, $\lambda_\mathrm{Bullock}$) indicators.
    \item Additionally, we have calibrated relations for the formation redshift, $z_\mathrm{form}$, as a function of our different indicators (Table \ref{tab:fits_zform}). With the exception of $\langle \tilde v_r \rangle$, these relations are tight and could be useful to understand the effects of selecting clusters based on their properties, and even to inform (semi-)analytic models (e.g., halo occupation distribution or similar methods).
\end{enumerate}

\item Looking at the correlations between instantaneous accretion rates and indicators, we obtain a clearer view on the timescale each indicator is informing about. Our results show how the indicators contain information about the accretion rates in different moments of the halo history. These range from instantaneous indicators ($\langle \tilde v_r \rangle$), to late-time accretion indicators ($\eta$ and $\chi$, sensitive to $z \in [0, 0.2]$ accretion rates), early-time accretion rate proxies ($f_\mathrm{sub}$, $\epsilon$, $\lambda_\mathrm{Bullock}$, around $z \simeq 0.3$), and finally $s_{200c,500c}$ provides the earliest information, being able to partially constrain the accretion rate up to $z \simeq 0.4$ when measured at $z=0$. A more detailed summary is presented in Table \ref{tab:summary}.

\item We have studied the characterisation of the diversity of MAHs with a couple of parameters ($z_\mathrm{form}$ and $\alpha_\mathrm{MAH}$) that can be defined with closed analytical forms from the $M(a)$ curves. These parameters can be understood in great correspondence (although non-linearly) to the two principal components of the MAHs, as extracted from the PCA algorithm reproducing the analysis of \citet{Wong_2012}.

\item  We have characterised the information brought by each assembly state indicator on both biparametric representations of the MAHs. While, on the $z_\mathrm{form} - \alpha_\mathrm{MAH}$ space the indicators are essentially incapable of constraining the episodicity parameter $\alpha_\mathrm{MAH}$, they can bring richer information in the $\mathrm{PC}_1 - \mathrm{PC}_2$ space. \\[-0.7em]

In this space, some indicators can be more informative about the formation time (e.g., $\lambda_\mathrm{Bullock}$), and others mainly inform about how episodic/bursty the accretion history is, while being essentially unrelated to the assembly time (e.g., $\Delta_r$). Most parameters (including our combined indicator, $\chi$), however, lie in a middle ground, thus providing a combination of information on these two evolutionary properties. This information is summarised in an indicator-by-indicator basis in Table \ref{tab:summary}.

\item Regardless of the trends obtained in this work, it is important noting how, while here we have mostly concentrated on the effect of the assembly state indicators on the ensemble properties of haloes, assessing the formation history of an individual galaxy cluster/group based on single properties is very difficult. This is the case due to the high intrinsic (cluster-to-cluster) scatter in the relations (e.g., $1 \sigma$ uncertainties of $\Delta z_\mathrm{form} \sim 0.2$ in Table \ref{tab:fits_zform}; see also, for instance, figure 8 of \citealp{Wong_2012}). \\[-0.8em]

This is, at least in part, due to the non-trivial evolution of these indicators. While all of them tend to increase with recent strong assembly activity, their particular evolution during smooth accretion episodes and mergers needs not be trivial. For instance, $\langle \tilde v_r \rangle$ is by definition insensitive to mergers during their pericentric passages, as well as (to a smaller extent) are $\Delta_r$ and $\epsilon$. Due to the intrinsic interest of this topic, we plan to address the relation of these quantities to mergers in a future, separate work.

\item When the same indicators mainly used in this work, which were derived from the DM distribution, are derived (with minimal adaptations) from the ICM gaseous distribution, again, even though there are moderate correlations, the high scatter makes it essentially impossible to relate the DM morphological and dynamical parameters to gas properties (or vice-versa) on a cluster-to-cluster basis. For example, in the best case, measuring a larger spin parameter $\lambda_\mathrm{Bullock}$ from the gas distribution only implies a larger rotation of DM in $\sim 75\%$ of the cases. This suggests that assessing dynamical state of a cluster in observations from single (ICM) morphological parameters may not be an accurate method, or rather the conclusions may not be directly extrapolated from the ICM to DM. This could only possibly be overcome by the (likely, non-linear) combination of many different indicators, what might additionally be redshift- and mass-dependent as hinted by \citetalias{Valles-Perez_2023}. 
\end{enumerate}

The results presented in this work add up to the existing bibliographic corpus in suggesting that dynamical state is not a single, monolithic property \citep[e.g.,][]{Haggar_2024}, and complement previous approaches in similar directions \citep[e.g.,][]{Wong_2012} by incorporating baryons (which are known to modify several of the parameters involved in this study) and looking at the constraining power of these parameters on the actual accretion rates, as well as the timing of the accretion phenomenology. 

Overall, we have shown how \textit{(i)} stacked MAHs are strongly dependent, even up to $z \simeq 4$, on selection of the clusters and groups based on the value of their indicators at $z=0$; \textit{(ii)} the different indicators respond to mass accretion in distinct moments through the assembly history (from instantaneous, to a $\tau_\mathrm{dyn}$ ago; and \textit{(iii)} each indicator points in different directions in the parameter space summarising the MAHs, implying that they may be able to constrain, not only the formation time, but also secondary properties (e.g., how smooth or bursty this accretion is).

Future work will need to deal with how this information, here obtained from the full three-dimensional description offered by simulations, is kept or hindered when obtained from observations. When dealing with observations, it will also be interesting to see in how far these theoretical measures of cluster assembly can be related to, e.g., X-ray and SZ morphology (e.g., \citealp{Zenteno_2020, Capalbo_2021}), or properties of the galaxy populations \citep{Berrier_2009}. Conceptually, this involves several steps that should be carefully considered: the relation between DM halo and ICM properties/galaxy distribution under varying physical models (including feedback prescriptions), the effects of projection, and the observational constraints of present and future instruments.  In forthcoming works of this series, we will concentrate on the effect of cluster assembly on the dynamical and thermodynamical properties of the ICM.

\begin{acknowledgements}
We thank the anonymous referee for their valuable feedback, which has contributed to improve the presentation of this work. This work has been supported by the Agencia Estatal de Investigación Española (AEI; grant PID2022-138855NB-C33), by the Ministerio de Ciencia e Innovación (MCIN) within the Plan de Recuperación, Transformación y Resiliencia del Gobierno de España through the project ASFAE/2022/001, with funding from European Union NextGenerationEU (PRTR-C17.I1), and by the Generalitat Valenciana (grant CIPROM/2022/49). DVP acknowledges support from Universitat de València through an Atracció de Talent fellowship. The simulation and part of the analysis has been carried out using the supercomputer Lluís Vives at the Servei d'Informàtica of the Universitat de València.

This research has been made possible by the following open-source projects: Numpy \citep{Numpy}, Scipy \citep{Scipy}, Matplotlib \citep{Matplotlib}, FFTW \citep{FFTW}, \texttt{emcee} \citep{Foreman-Mackey_2013}.
\end{acknowledgements}
\bibliographystyle{aa} 
\bibliography{accretion-1}

\clearpage
\appendix

\section{Merger trees and MAHs}
\label{s:app.mtree}

As discussed in Sect. \ref{s:results.MAHs} (Fig. \ref{fig:MAH_mass}), the ensemble MAHs for our three mass subsamples depart significantly from the analytic results of \citet{Correa_2015} and, albeit on a smaller level, from the semi-analytic results of \citet{vandenBosch_2014}. This is better quantified in Table \ref{tab:errors_mah}, where in the top row of each group we give the mean relative errors (MRE) of our MAHs with respect to their tabulated ones, for the three mass ranges. Although \citet{vandenBosch_2014} MAHs reduce the error up to a factor of 2, it is still above $25\%$ for some mass ranges.

In this work, we define the main branch of the merger tree by picking, at each step backwards in time, the main progenitor as the unique halo at the previous snapshot which:

\begin{itemize}
    \item Contains, amongst its particles, the most-gravitationally bound particle of the descendant halo.
    \item Its most-gravitationally bound particle is within the descendant particles.
\end{itemize}

\begin{figure}
    \centering
    \includegraphics[width=\linewidth]{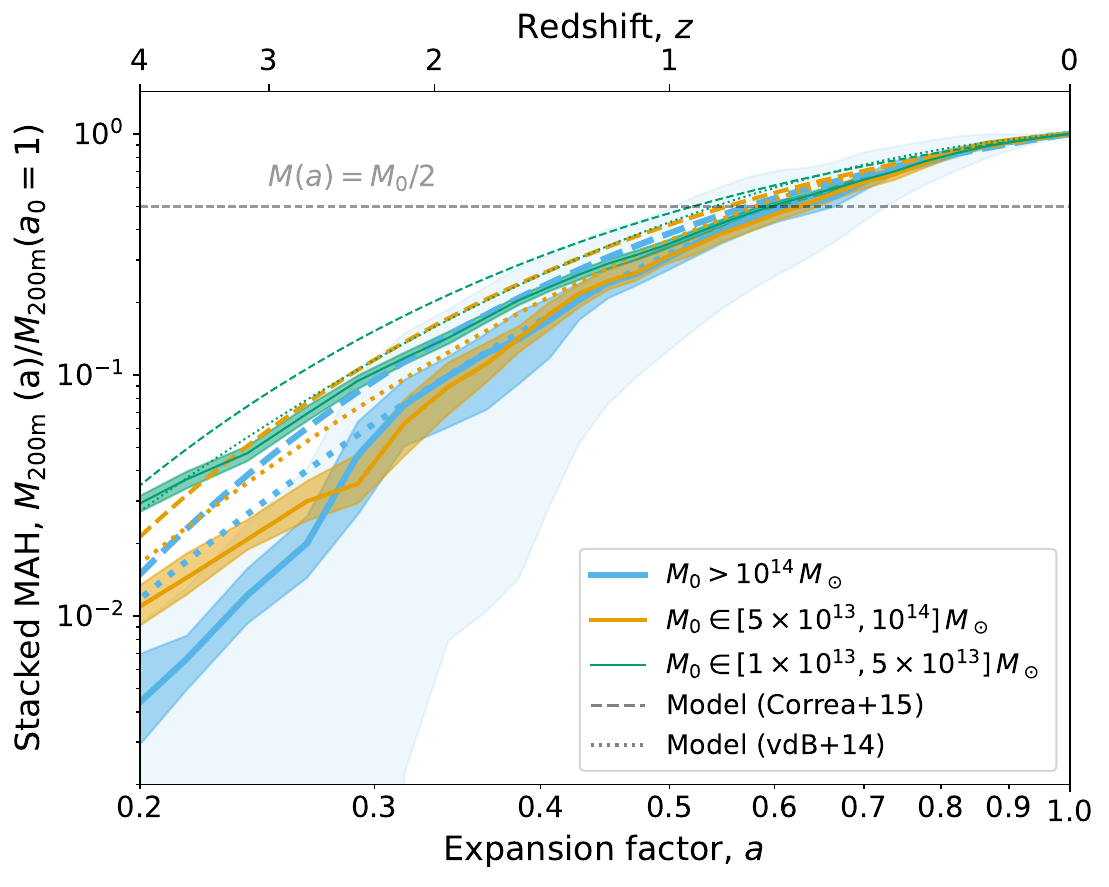}
    \caption{Equivalent to Fig. \ref{fig:MAH_mass} of the main text, with the main branch of the merger tree having been set by choosing the most massive progenitor at each timestep.}
    \label{fig:MAH_mass_appendix}
\end{figure}

This criterion (hereon, the \textit{most-bound} method) can sensibly depart from the one based on picking the most-massive progenitor at each snapshot, used both in \citet{vandenBosch_2014} and \citet{Correa_2015} (the \textit{most-massive} method). In Fig. \ref{fig:MAH_mass_appendix} we show an equivalent figure to Fig. \ref{fig:MAH_mass}, where the main branch of the merger tree has been picked by the most-massive method. In this case, we see how our recovered MAHs for the most massive subsample are fully consistent with \citeauthor{vandenBosch_2014}'s (\citeyear{vandenBosch_2014}) within the $1 \sigma$ mean error, and the discrepancy at lower masses is greatly reduced (see also the bottom row in each group of Table \ref{tab:errors_mah}).

We have also fitted our original MAHs and the ones obtained with the most-massive method to the functional form suggested by \citet{McBride_2009}, which is the one used by \citet{Correa_2015} as well,

\begin{equation}
    M(z) = M_0 (1+z)^\alpha e^{\beta z},
    \label{eq:correa15}
\end{equation}

\noindent which describes an exponential phase at high redshift (${\Gamma |_{z \to \infty} \to - \beta / a}$) and a more relaxed, power-law regime at low redshift ($\Gamma |_{z \to 0} \to - (\alpha + \beta)$). The fits have been performed using a Markov-chain Montecarlo ensemble sampler (implemented in the \texttt{emcee} package, \citealp{Foreman-Mackey_2013}), with flat priors and a $\chi^2$ log-likelihood, to recover the whole posterior distribution of the fit parameters. The fit results are summarised in Table \ref{tab:fits_correa}. 

While confidence intervals are generally large and may contain the reference values in both cases, both the coefficients of determination (larger for the most-massive method) and the more continuous variation of the parameters with mass clearly suggest a better compatibility of this latter method with the functional form of \citet{McBride_2009}.

Notably, we observe that these two criteria tend to converge when selecting, from the whole sample, the objects with the narrowest MAHs (the least quotient between $z=4$ and $z=0$ masses). This suggests that the differences between methods appear mainly in periods of strong assembly activity (e.g., major mergers), which is precisely when the largest discrepancies among merger trees arise \citep{Behroozi_2015}.

However, it is important stressing that there is nothing incorrect with either choice of main branch of the merger tree. While in the main text we have chosen to follow the most-bound method because of its sensible physical interpretation, this section proves that our simulation results can also reproduce the previous literature. 

\begin{table}
    \centering
    \small
    \caption{Percentual mean relative error (MRE) between the tabulated MAHs of \citet{vandenBosch_2014} and \citet{Correa_2015}, and the ones obtained from our simulation. For each group of two rows, the upper row corresponds to the \textit{most-bound} method (used in this work), while the bottom row has been obtained by defining the main branch of the merger tree using the \textit{most-massive} progenitor.}
    \begin{tabular}{cc|c|c}
    & & \multicolumn{2}{c}{MRE ($\%$)} \\
         \multicolumn{2}{c|}{Mass range (at $z=0$)}& Correa+15 & vdB+14  \\ \hline \hline
         \multirow{ 2}{*}{$>10^{14} M_\odot$} & most-bound &  50.2 & 23.3  \\
          & most-massive & 29.3 & 9.1 \\ \hline
          \multirow{ 2}{*}{($5 \times 10^{13} - 10^{14}) M_\odot$} & most-bound & 51.1  & 26.4  \\
          & most-massive & 30.0 & 10.7 \\ \hline
          \multirow{ 2}{*}{($1 \times 10^{13} - 5 \times 10^{13}) M_\odot$} & most-bound & 27.1 & 15.0 \\
          & most-massive & 29.7 & 16.4 \\ 
    \end{tabular}
    
    \label{tab:errors_mah}
\end{table}

\begin{table}
    \centering
    \small
    \caption{Fit parameters ($\alpha$ and $\beta$) for our three mass ranges to the functional form in Eq. (\ref{eq:correa15}). These results are shown for the merger trees used in the paper results (top block), and for the merger trees computed from the most massive progenitors at each snapshot (bottom block). The reference values for $\alpha$ and $\beta$ are obtained from \citet{Correa_2015} at the median mass in each subsample. Within each pair of rows, in the rightmost column, the top row shows the coefficient of determination, while the bottom row shows the mean relative error between best-fit and \citet{Correa_2015}. }
    \begin{tabular}{c|c|c|c}
        \hline \hline
        \multicolumn{4}{c}{Main branch of the merger tree: most bound} \\ \hline
        Mass range (at $z=0$) & $\alpha$ & $\beta$ & $R_\mathrm{fit}^2$ / MRE \\ \hline
    $M_\mathrm{vir} > 10^{14} M_\odot$ & $0.66^{+1.13}_{-0.44}$ & $-1.58^{+0.24}_{-0.79}$ & 0.93 \\
    ... Reference & $0.36$ & $-1.19$ & $50\%$ \\ \hline
    $5 \times 10^{13} M_\odot < M_\mathrm{vir} < 10^{14} M_\odot$ & $0.73^{+1.26}_{-0.46}$ & $-1.60^{+0.27}_{-0.91}$ & $0.70$ \\
    ... Reference & $0.32$ & $-1.09$ & $51\%$  \\ \hline
    $10^{13} M_\odot < M_\mathrm{vir} < 5 \times 10^{13} M_\odot$ & $-0.51^{+0.90}_{-0.26}$ & $-0.66^{+0.10}_{-0.55}$ & $0.99$ \\
    ... Reference & $0.28$ & $-0.95$ & $27\%$ \\ \hline\hline
    \multicolumn{4}{c}{Main branch of the merger tree: most massive} \\ \hline
    Mass range (at $z=0$) & $\alpha$ & $\beta$ & $R_\mathrm{fit}^2$ / MRE \\ \hline
    $M_\mathrm{vir} > 10^{14} M_\odot$ & $0.10^{+0.87}_{-0.25}$ & $-1.18^{+0.11}_{-0.56}$ & 0.99 \\
    ... Reference & $0.36$ & $-1.19$ & $29\%$ \\ \hline
    $5 \times 10^{13} M_\odot < M_\mathrm{vir} < 10^{14} M_\odot$ & $0.10^{+1.21}_{-0.30}$ & $-1.14^{+0.16}_{-0.84}$ & 0.97 \\
    ... Reference & $0.32$ & $-1.09$ & $30\%$ \\ \hline
    $10^{13} M_\odot < M_\mathrm{vir} < 5\times 10^{13} M_\odot$ & $-0.01^{+0.94}_{-0.31}$ & $-0.96^{+0.14}_{-0.62}$ & 0.99 \\
    ... Reference & $0.28$ & $-0.95$ & $30\%$  \\ \hline \hline
    \end{tabular}
    \label{tab:fits_correa}
\end{table}

\section{PCA of MAHs}
\label{s:app.PCA}

In Sect. \ref{s:results.MAHs_biparametric.biparametric}, we have compared our biparametric description of the MAHs to their first two principal components, in order to assess in how far our analytically-defined parameters reproduce a large fraction of the diversity of this family of curves. In this Appendix, we extend on the description of the Principal Component Analysis (PCA) performed here, and compare it to earlier works. 

\subsection{PCA methodology and results}
\label{s:app.PCA.method}

\begin{figure*}[h]
    \centering
    \includegraphics[width=\linewidth]{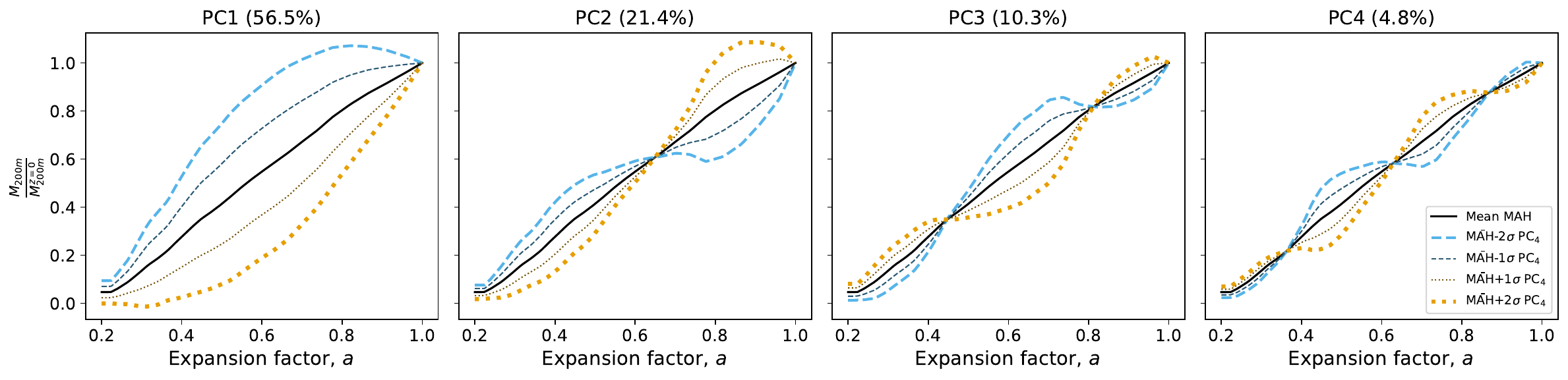}
    \caption{The first four principal components of our MAHs (from left to right, in decreasing order of explained variance, stated in the title of each panel). Within each panel, the black line represents the mean MAH of the sample. The different additional curves correspond to $\pm 1 \sigma$ and $\pm 2 \sigma$ of the values of the corresponding principal component, reflecting how each component is producing changes in the MAHs of our sample.}
    \label{fig:PCA_real_appendix}
\end{figure*}

PCA \citep{Pearson_1901, Hotelling_1936} is a relatively simple method to extract the linearly uncorrelated components of a certain dataset (i.e., the eigenvectors of the correlation matrix), in such a way that one can keep the ones accounting for the largest fractions of variance to reduce the dimensionality of the input space. \citet{Wong_2012} applied this methodology to the characterisation of the assembly of their DM haloes with their first two principal components. Given the different masses and selection methods in our sample with respect to the aforementioned reference, instead of using their PCs, we perform the PCA on our data and compare to their results.

In practical terms, we keep all haloes that can be tracked down to at least $z_\mathrm{start}=4$ ($\sim 50\%$ of our total sample), and we resample the MAH of the $j$-th halo, $\mathcal{M}_j(a) \equiv M_j(a)/M_{j0}$ onto $N_a = 101$ points equally spaced in the expansion factor $a$. In this way, we obtain $N_\mathrm{haloes}=187$ MAHs, described as vectors in $\mathbb{R}^{101}$. Since PCA is sensitive to the relative scaling of the components, we standardise the input data, such that

\begin{equation}
    \widetilde{\mathcal{M}}_j (a_i) = \frac{\mathcal{M}_j (a_i) - \langle \mathcal{M}(a_i) \rangle}{\sigma \left( \mathcal{M}(a_i)\right)},
\end{equation}

\noindent where the average ($\langle \cdot \rangle$) and standard deviation ($\sigma(\cdot)$) are performed over the $N_\mathrm{haloes}$ independently at each $a_i$, so that the set of $\left\{\widetilde{\mathcal{M}}_j (a_i) \right\}_{j=1}^{N_\mathrm{haloes}}$ has null mean and unit variance. Finally, the eigenvectors of the correlation matrix,

\begin{equation}
    \rho_{ij} = \frac{\langle \widetilde{\mathcal{M}}(a_i) \widetilde{\mathcal{M}}(a_j) \rangle}{\sigma \left( \widetilde{\mathcal{M}}(a_i)\right) \sigma \left( \widetilde{\mathcal{M}}(a_j)\right)},
\end{equation}

\noindent yield the PCs, $\left\{ \widehat{\text{PC}}_i \right\}_{i=1}^{101}$, while their eigenvalues inform about the fraction of variance in this $\mathbb{R}^{101}$-dimensional distribution of MAHs explained by the corresponding PC.

Using this methodology, Fig. \ref{fig:PCA_real_appendix} shows graphically the first four principal components of the data, by adding $\pm 1 \sigma$ and $\pm 2 \sigma$ to the mean MAH, $\sigma$ being the standard deviation of the values of the corresponding component of the linear decomposition, $\mathrm{PC}^i_j = \widetilde{\mathcal{M}}_j(a) \cdot \widehat{\text{PC}}_i$ (where $\cdot$ denotes the dot product in $\mathbb{R}^{101}$).

The results, shown from left to right for the first four principal components in decreasing order of variance explained, are qualitatively in broad agreement with \citet{Wong_2012} (their figure 2, showing the first two PCs); namely, $\text{PC}_1$ showing two nodes at opposite ends\footnote{At $a_0=1$, we naturally encounter a node since $\mathcal{M}_j(a_0=1)=1$ for all $j$ by construction. At the high-redshift end, however, we find an \textit{approximate} node: while here not all values of $\mathcal{M}_j(a \approx 0.2)$ are equal, they typically lie in the interval $[0.01, 0.1]$, thus much smaller than $1$.} and a central antinode. For $\text{PC}_2$, an additional central node appears and therefore the second component describing the most variability in the set of MAHs explains a behaviour with two antinodes. The situation is reproduced for subsequent principal components, $\text{PC}_3$ ($\text{PC}_4$) describing $\mathcal{M}(a)$ curves with 3 (4) antinodes. The origin of this rather peculiar behaviour is explored in more detail in Sect. \ref{s:app.PCA.interpretation}.

We note, however, that quantitative differences between these results and the ones by \citet{Wong_2012} emerge upon closer inspection. In particular, regarding $\text{PC}_2$, while their node is located at $a \simeq 0.45$ (and hence this component is related to the instantaneous accretion rate especially at this particular epoch), ours is located at $a \simeq 0.65$. This could be well explained by the different mass interval of the two studies, together with a different strategy for building the merger tree.

\subsection{Interpretation of the principal components}
\label{s:app.PCA.interpretation}

In order to understand why these PCs emerge from our data, we have constructed a set of \textit{mock} MAHs, which are not extracted from any simulation data but only aim to reproduce the overall shape of these curves, instead.

\begin{figure*}
    \centering
    \includegraphics[width=\linewidth]{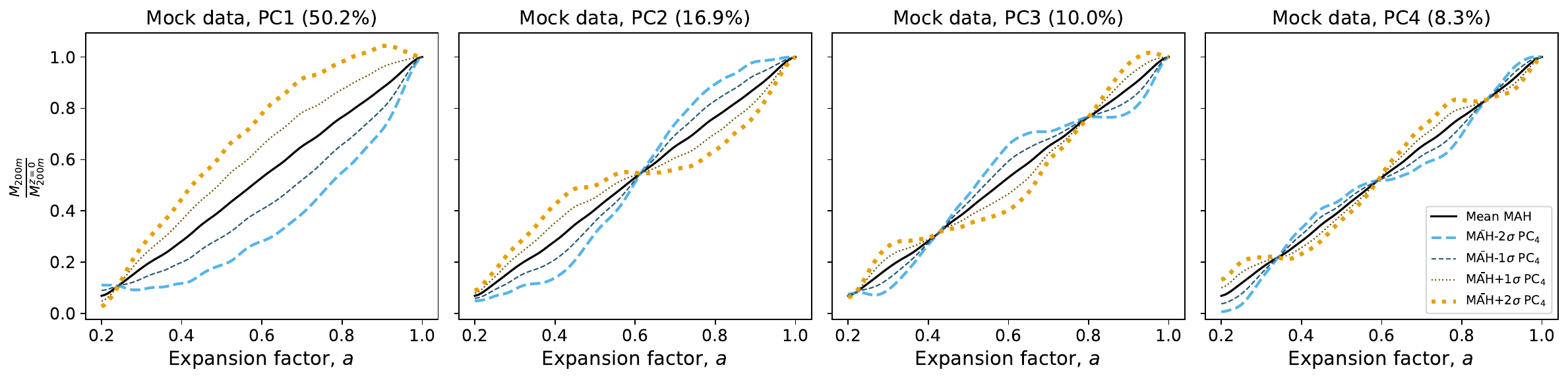}
    \caption{Equivalent to Fig. \ref{fig:PCA_real_appendix}, but for the mock MAHs. All figure elements are equivalent to the previous one.}
    \label{fig:PCA_mock_appendix}
\end{figure*}

\begin{figure}[h]
    \centering
    \includegraphics[width=\linewidth]{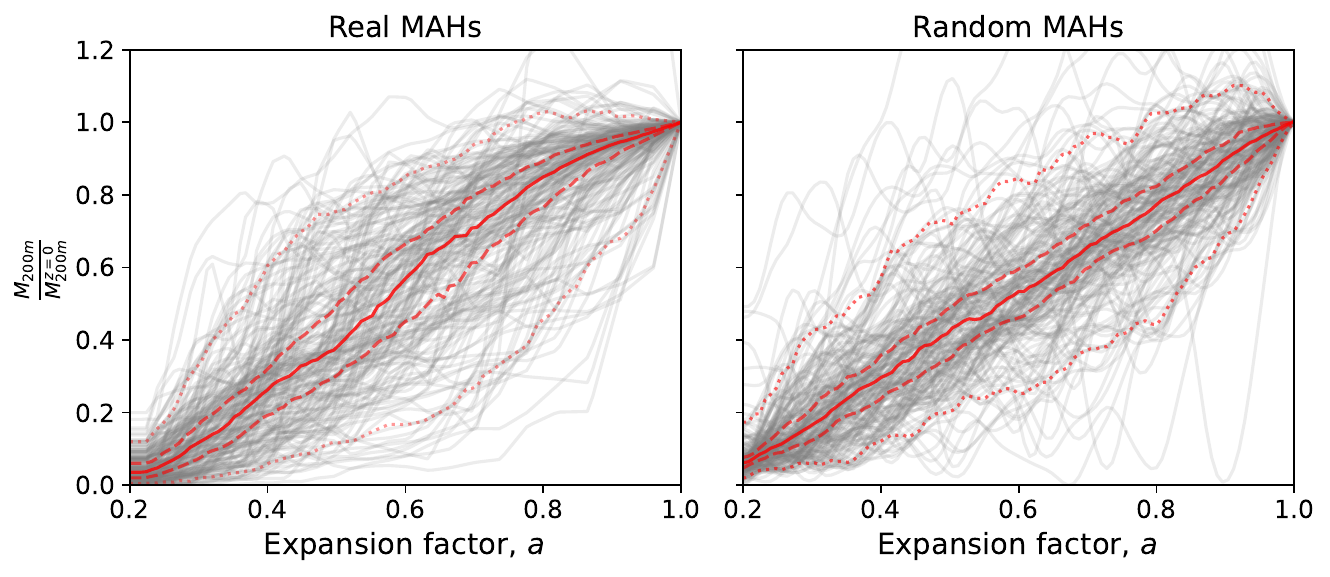}
    \caption{Comparison between the actual MAHs in our sample (left-hand side panel) and the mock ones generated to investigate the origin of their PCA (right-hand side panel). Here, gray lines are individual MAHs. The solid, red line is the mean MAH, while the dashed (dotted) lines indicate $32-68$ ($5-95$) percentiles.}
    \label{fig:mock_MAHs}
\end{figure}

Each mock MAH is generated as follows:

\begin{enumerate}
    \item We generate $N_a$ random numbers $u_i \sim \mathcal{N}(0,1) + 0.3$, where $\mathcal{N}(0,1)$ is a standardised normal distribution. This allows for the MAHs to locally decrease (as it may happen in actual haloes) but to present an overall increasing trend.
    \item We obtain the cumulative sum, $U_i = \sum_{j=1}^i u_j$, and then smooth it with a Gaussian filter with $\sigma = 2$ points to get rid of small-scale oscillations.
    \item In order to reproduce the variability at the high-redshift end, we finally define the normalised MAH as:
\end{enumerate}
\begin{equation}
    \mathcal{M}(a_i) = \max(0.05+x, 0) + \min(0.95-x, 1) \frac{U_i - \min_j U_j}{U_{N_a} - \min_j U_j}
\end{equation}

\noindent where $x$ is yet another random number drawn from the $\mathcal{N}(0,0.02)$ distribution. Empirically, we find that this heuristics produces curves with a shape remarkably similar to the real MAHs, as shown in Fig. \ref{fig:mock_MAHs}.

The PCA of the mock MAHs is shown in Fig. \ref{fig:PCA_mock_appendix}, in a completely equivalent way to Fig. \ref{fig:PCA_real_appendix} above for the real MAHs. Besides a slightly less smooth morphology in some cases, which can be attributed to the high-frequency noise in the mock MAHs, the mock data appears to distribute in a very similar way to the real one. Note that the only requirements imposed on these curves are:

\begin{itemize}
    \item Having a fixed extreme at $a = 1$ ($\mathcal{M} = 1$), and having an \textit{almost-fixed} extreme at $a \simeq 0.2$ ($\mathcal{M} \ll 1$).
    \item Having an overall increasing trend, although decreasing intervals are allowed.
\end{itemize}

Therefore, it appears that these principal components, obtained by \citet{Wong_2012} and which we reproduce in this work, which are conceptually similar to the harmonic modes in a one-dimensional system with the two fix boundary conditions, emerge primarily as a consequence of these limiting behaviours of the MAH curves, namely $\lim_{a \to 1} \mathcal{M}(a) = 1$ and $\lim_{a \to a_\mathrm{start}} \mathcal{M}(a) \ll 1$, and are probably not as much related to specific details about the whole span of the MAH (i.e., they do not seem to be related to possible physical properties such as, e.g., typical timescale between mergers).

This does not detract from the fact that the first two principal components are capable of explaining $77\%$ ($83\%$ in \citeauthor{Wong_2012}'s \citeyear{Wong_2012} case) of the variance in the (standardised) MAHs, implying that they constitute an idoneous base where to summarize their complex diversity. Indeed, as seen through Sect. \ref{s:results.MAHs_biparametric} and especially in Fig. \ref{fig:variation_directions}, the different indicators involved in this work seem to inform about different combinations of these two principal components, while other perhaps more physically motivated definitions aiming to summarise the MAHs do not allow to extract as much information.

\end{document}